\documentclass[10pt,journal]{IEEEtran}

\usepackage[english]{babel}
\usepackage{amsmath}      % Defines certain mathematical symbols
\usepackage{mathbbol}     % used for the vector of ones
\usepackage{verbatim}     %to use the \begin{comment}
\usepackage{amsthm}
\usepackage{amssymb}
\usepackage{amsfonts}
\usepackage{url}               % To typeset email addresses and URLs
\usepackage[pdftex]{graphicx}
\DeclareGraphicsExtensions{.eps}

\usepackage{pdfpages}

\usepackage{subfigure} % for the case of figure 4, varying M

\usepackage{hyperref}
\usepackage{indentfirst}
\usepackage{cite}
\usepackage{setspace}
\usepackage{units}
\usepackage[short]{optidef}

\begin{document}
%\title{Massive Wireless Energy Transfer with Limited Channel Knowledge}
%\title{Wireless Energy Transfer with Limited CSI\\for Massive MTC}
\title{Massive Wireless Energy Transfer\\ with Statistical CSI Beamforming}
%\title{User Fairness for Wireless Energy Transfer with Massive MIMO with Limited Channel Knowledge}

\author{Francisco~A.~Monteiro,~\IEEEmembership{Member,~IEEE,}
        Onel L. A. López,~\IEEEmembership{Member,~IEEE,}
        Hirley Alves,~\IEEEmembership{Member,~IEEE}
\thanks{This work was supported by the Academy of Finland (Aka) (Grants n.307492, n.318927 (6Genesis Flagship), n.319008 (EE-IoT)). The work of Francisco Monteiro was supported by a joint scholarship from Fundación Carolina and Fundación Endesa (both from Spain) and funded by FCT/MCTES (Portugal) through national funds and when applicable co-funded EU funds under the project UIDB/50008/2020.}
\thanks{F. A. Monteiro is with Instituto de Telecomunicações, and ISCTE - Instituto Universitário de Lisboa, Portugal, e-mail: \{francisco.monteiro@lx.it.pt\}.}
\thanks{Onel L. A. L\'opez and Hirley Alves are  with  the Centre for Wireless Communications, University of Oulu, Finland, e-mails: \{Onel.AlcarazLopez@oulu.fi, Hirley.Alves@oulu.fi\}.}% <-this % stops a space

%\thanks{This work was funded by FCT (Foundation for Science and Technology) and Instituto de Telecomunica\c{c}\~{o}es through national funds, and when applicable co-funded EU funds, under the project UIDB/EEA/50008/2020}
} 

\maketitle
\begin{abstract}
Wireless energy transfer (WET) is a promising solution to enable massive machine-type communications (mMTC) with low-complexity and low-powered wireless devices. Given the energy restrictions of the devices, instant channel state information at the transmitter (CSIT) is not expected to be available in practical WET-enabled mMTC. However, because it is common that the terminals appear spatially clustered, some degree of spatial correlation between their channels to the base station (BS) is expected to occur. The paper considers a massive antenna array at the BS for WET that only has access to i) the first and second order statistics of the Rician channel component of the  multiple-input multiple-output (MIMO) channel and also to ii) the line-of-sight MIMO component. The optimal precoding scheme that maximizes the total energy available to the single-antenna devices is derived considering a continuous alphabet for the precoders, permitting any modulated or deterministic waveform. This may lead to some devices in the clusters being assigned a low fraction of the total available power in the cluster, creating a rather uneven situation among them. Consequently, a fairness criterion is introduced, imposing a minimum amount of power allocated to the terminals. A piece-wise linear harvesting circuit is considered at the terminals, with both saturation and a minimum sensitivity, and a constrained version of the precoder is also proposed by solving a non-linear programming problem. A paramount benefit of the constrained precoder is the encompassment of fairness in the power allocation to the different clusters. Moreover, given the polynomial complexity increase of the proposed unconstrained precoder, and the observed linear gain of the system's available sum-power with an increasing number of antennas at the ULA, the use of massive antenna arrays is desirable.

\end{abstract}

\section{Introduction}

The internet of things (IoT) presupposes a large number of low-complexity devices, most of them wireless, and a number of them placed in hardly accessible locations (e.g., for infrastructure monitoring). Even though most of these terminals are not supposed to transmit continuously, they are supposed to be able to operate for long periods of time without batteries replacement. One way of achieving that is by having devices that can harvest energy from incoming electromagnetic radiation. Much work has been recently done on simultaneous wireless information and power transmission (SWIPT) \cite{added_SWIPT}, wireless-powered communication networks (WPCN), in which terminals harvest energy during a first time slot and use it to power the transmission of information in a second time slot\cite{added_a10}, and also on systems dedicated to wireless energy transfer (WET) \textit{per se} \cite{Zhang2016, WETmag,Alsaba2018,added_R1,Added_b_Zhang16,lopez_statistical_2019, kashyap_feasibility_2016, WCL,Added_c_Heath,added_a33}.
WET systems differ from the former two by only focusing on the transfer of energy without spending resources on any type of data transfer.

As it is thoroughly described in \cite{added_a36}, the idea of WET is as old as radio transmission but, until quite recently, the idea typically involved high transmission power and very large antennas at both ends of a link, and often aimed at very long distances (e.g., satellite-to-Earth links). Using low-power WET to power devices over a few meters (tens or even hundreds of meters) is a quite recent endeavor.
The WET approach is easier to deploy massively given that the energy source, or power beacon (also still called base station (BS) for legacy reasons), does not need to have any connection to a core data network apart from the connection to the electrical grid. Note that the  WET component of a system is often the one responsible for performance bottleneck in practical WPCN and SWIPT systems, and for that reason it is worthy of dedicated study and consideration of different optimization approaches. Other examples of systems that may also need to rely on WET are backscatter communications \cite{liu_next_2019}, and the new long-distance tags based on quantum tunneling radio positioning\cite{quantum_RFID_2020}.

In \cite{Alsaba2018}, one can find a survey of beamforming techniques for SWIPT, WPCN, and WET systems, and also a list of some techniques to easy the enormous task of acquiring channel state information at the transmitter (CSIT), on which those beamforming techniques depend on.
However, instantaneous CSIT availability is an unrealistic assumption in the context of mMTC because not only the devices are very energy-constrained, but also due to the immense number of pilots the BS would have to deal with in the uplink. Some steps forward to do away with instantaneous CSIT have been recently taken in \cite{Clerckx_Kim_2018}, where fading is artificially accentuated such that its benefits can be further exploited even if CSIT is entirely absent, and in \cite{lopez_statistical_2019}, where a statistical analysis of the harvested energy in a WET setup assuming no CSIT is presented. In \cite{lopez_statistical_2019}, the authors assessed WET with single antenna terminals and a multiple antenna transmit array at the BS, and compared i) single antenna transmission from the BS, ii) equal power transmission from all the BS's antenna elements, dubbed as all antenna at once (AA) scheme, and iii) the switching antennas (SA) scheme, where only one antenna is active at a time and the position of the active one runs through all the antennas in the array during each block of coherent fading. Their channel model had all devices experiencing a Rician fading channel and distributed around the BS without any particular distinction between devices.

A great part of the literature on WET-enabled systems (WET only or SWIPT) assumes that full instantaneous channel knowledge is available at the BS, e.g., \cite{kashyap_feasibility_2016,TimKrik2014,Thudugalage.2016,Moraes.2017,Clerckx.2018,Choi.2018}, or even simultaneously available at an intermediate intelligent reflective surface \cite{Lyu.2020}.
Nevertheless, imperfect CSIT has also been considered in a a number of works \cite{Added_b_Zhang16,added_R1,Added_c_Heath,added_a33,kashyap_feasibility_2016}. In general, acquiring a more accurate information about the channel requires more energy expenditure, and consequently there is a trade-off between the energy that is harvested by the terminals and the energy spent feed-backing CSI. The works in \cite{added_R1} and \cite{kashyap_feasibility_2016} considered point-to-point WET, with a massive antenna array wirelessly powering only one EH multi-antenna terminal. In \cite{added_R1} the EH device has multiple antennas and the aim was the maximization of the net harvested energy by optimizing the training phase and the WET phase. In \cite{kashyap_feasibility_2016} the authors considered a multi-input single-output (MISO) system, i.e., with a single-antenna
EH terminal, and derived expressions for the outage probability of EH, defining outage when the harvested energy during a block is less than the sum of the energy the sensor spent in transmitting the CSI and the energy spent processing its tasks. In both cases a Rician channel model was considered and the channel matrix is written as the sum of a known matrix (representing the partial CSIT) with a unknown remaining term for a Rician channel, which is the most common way of incorporating partial CSIT in the system models. The work in \cite{added_R2} also considers massive antenna array at the BS but rather a multi-user (MU) system with single-antenna EH terminals, and formulates partial CSIT in the same way. Similarly, in the context of SWIPT, the lack of full-CSIT has also been studied in the same manner, e.g., in \cite{added_a30} a MU-SWIPT system is considered with single-antenna terminals.

This paper considers partial CSIT, formulated in a distinct manner, such that the partial knowledge comprises of \textit{statistical} knowledge about the channel, which has been argued to be beneficial for MIMO WET systems in \cite{WETmag}. Note that in open-loop WET systems (i.e., CSIT-free), which are oblivious of the channel state and operate without any CSIT \cite{IoTJ_2021,lopez_statistical_2019,Clerckx_Kim_2018}, only a statistical analysis of the harvested energy is possible, and a sizeable statistical analysis of such open-loop systems currently exists. In \cite{WCL}, a fair beamforming scheme for WET  that only makes use of the channel's first order statistics (i.e., the channels' mean) was proposed, also under Rician fading. The present paper proposes reduced-complexity precoding that also does not require instantaneous full-CSIT but only the first and second order statistics of the channel, considering that both the channel correlation and the LOS components vary slowly.

\begin{figure}[t]
\begin{center}
   	\includegraphics[width=0.8\columnwidth, clip=true, draft=false]{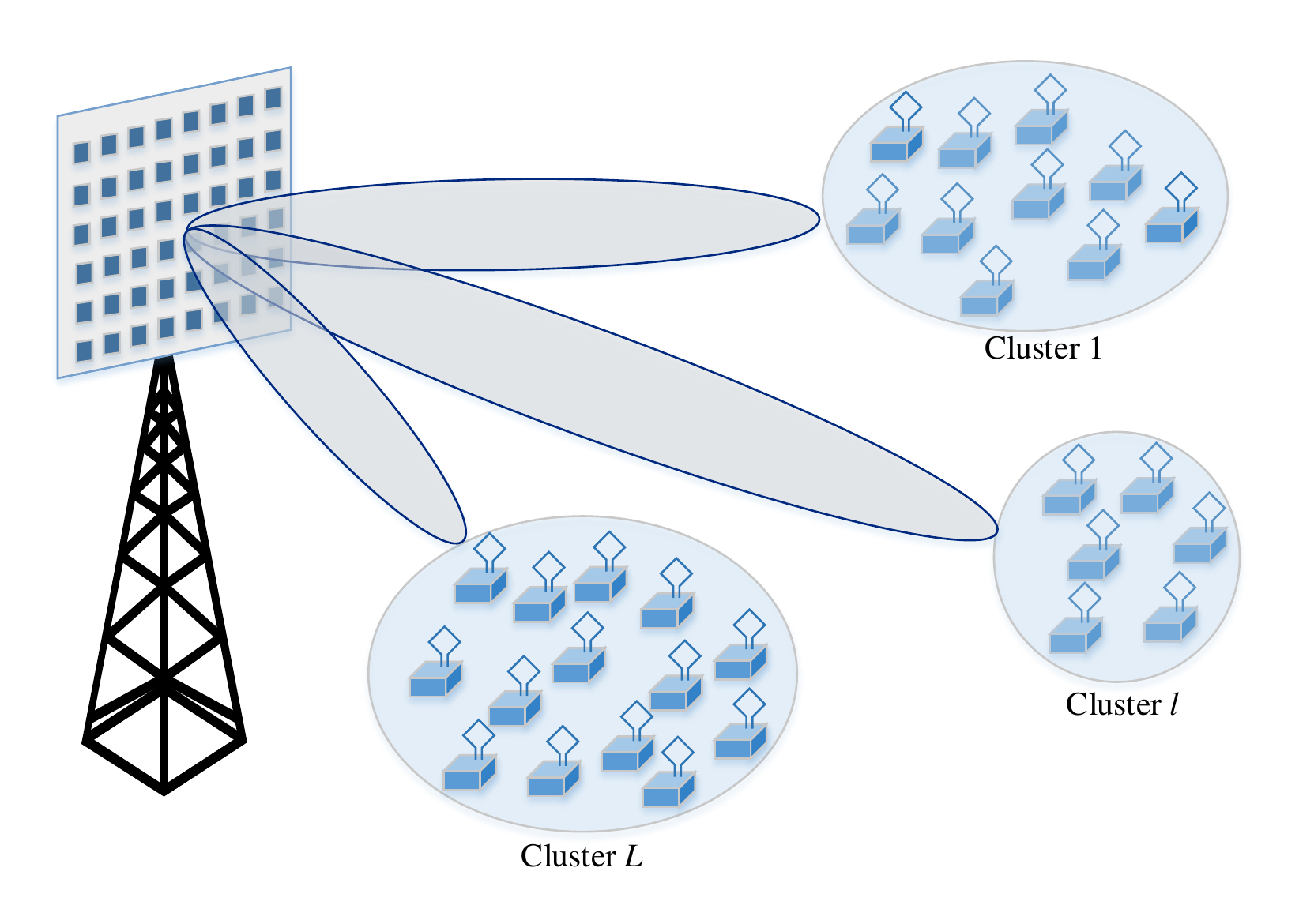}
\end{center}
   	\caption[ccc]{WET system model with $L$ clusters, each one with $K$ MTC terminals, powered by a BS with $M$-antennas applying MIMO precoding with limited CSIT.}
\label{Fig:system_setup}
\end{figure}

In many practical scenarios the terminals are physically clustered (as in Fig. \ref{Fig:system_setup}) and thus, besides sharing the same large-scale gain (pathloss) from the BS to the terminals, they also share a common slow-fading component, which can be characterized by its second order statistics, i.e., the covariance of the channel matrix.
This paper proposes to take advantage of that partial CSIT, which can be estimated with pilots transmitted from the terminals to the BS at a low rate.

The Karhunen–Loève (KL) channel representation allows one to use beamforming techniques reminiscent of the initial ideas for low-CSIT MIMO precoding with reduced CSI feedback \cite{simeone_combined_2003}.
Similar considerations about the correlation between the channels experienced by nearby terminals lead to a proposal combining massive MU-MIMO, MIMO precoding, and user clustering \cite{ding_design_2016}. In that scenario, beamforming is performed for the different clusters of users, and then the intra-cluster users use non-orthogonal multiple access (NOMA) in the power domain to detect their own signals.

It should be noted that the proposed system is a MU-MIMO setup in which each cluster assumes the role of a virtual single-antenna terminal, representing all the terminals sharing a common second order statistics and a similar line-of-sight (LOS) model. This is in contrast with \cite{kashyap_feasibility_2016}, where a MISO system is considered for WET in order to highlight the benefits of using a massive antenna array.

The energy beams can be adapted to focus onto each cluster and, because interference is not a foe in the context of WET, the optimal precoder can be designed by maximizing the total system's sum-power available to be harvested at the terminals' antenna in all clusters. However, in doing so, an unfair partial sum-power allocation among different clusters may happen and it is shown that by framing the problem as a constrained numerical optimization, fairness can be achieved among clusters. A different approach to introduce fairness in a WET MU-MIMO scenario was taken in \cite{Thudugalage.2016}, where the authors proposed designing a precoder which takes in consideration fairness by applying a max-min criterion to maximize the network's user with the weakest channel conditions, while considering full-CSIT and Rayleigh fading. In that work, the authors considered a different pathloss for each device, given that they are not considered to be clustered around each other. To the best of our knowledge, a WET system model considering clusters of terminals has only appeared in the very recent papers \cite{IoTJ_2021,IoTJ_2021B}, but the fact that they are considering clusters has no implications for precoding in these works, which assess open-loop techniques such as all antennas at once (AA) and switching antennas (SA).

One focuses on the available power to be harvested rather than the harvested energy after the energy harvesting (EH) circuit in order to offer more general results. Nevertheless, the specifications of a particular piece-wise linear EH circuit (to be detailed in section \ref{Sec:system_model}) are considered in this work in order to tune the system to a particular useful power range.

Besides the limitation due to the high sensitivity of the EH circuits (e.g., around $-10$ dBm, compared to around $-60$ dBm when the goal is information transmission), there is also the downside of the inefficiency of the voltage multipliers based in diodes chiefly responsible for converting the radio-frequency energy (RF) into direct current (DC) voltages for both immediate use and storage, as highlighted in \cite{added_a07}, and thoroughly described in \cite{added_survey}. A description of different antennas designs for the EH terminals and a compilation of EH circuits characteristics, such as frequency range and power conversion efficiency and be found in \cite{added_EHcircuits}.

The harvested DC power is not only a non-linear function of the RF power at the antenna for a given transmitted waveform, but also a function of the waveform itself \cite{Added_Clerckx21}. For an easier approach and analysis of energy beamforming, previous works have assumed a fully-linear conversion, i.e., considering a linear function for the RD-DC conversion, and not even considering the sensitivity and saturation effects of the circuits, e.g., \cite{added_a12}. The same simplification was made to design and analyze systems with a WET phase followed by an information transfer phase from sensors \cite{added_a13, added_a14,added_a15,added_a16}, and for MU-MIMO WET with beamforming \cite{Thudugalage.2016}, as in the scenario considered in the present paper.
In fact, this simple linear model can be traced back to \cite{added_Diode}. Nevertheless, more realistic, yet more challenging, non-linear RF-DC conversion models have been considered in the literature, ranging from characterizing the non-linearity by a sigmoid function  \cite{added_a17}, to assuming a general RF-DC transfer function that only needs to be monotone and increasing in \cite{added_a18}. A model with saturation-only was proposed in \cite{added_a31,added_a33}, and also used in \cite{added_a34} for beamforming in SWIPT.

A significant step forward was taken in \cite{added_a35}, by characterizing the RF-DC conversion while keeping analytical tractability. In doing that, the authors have found that the (unmodulated) single sinewave was not the optimal waveform for WET when the non-linearity is considered and proposed deterministic multisine waveforms. These waveforms, which have a higher PAPR (peak-to-average power ratio) when the number of sines increases, eventually deliver a higher output DC power that scales linearly with the number of sine waves even without CSIT, as long as the the channel fading is flat.  Under frequency-selective fading the linear increase is also attainable as long as CSIT is present \cite{added_a35}.
Interestingly, the benefits of using multisine waveforms for WET had been previously hinted in \cite{added_multisine}, where the authors proposed using multi-sinewaves to better match the Taylor expansion of the diode's non-linear function, and also for \cite{added_RFID} for RFID tags, in which the authors proposed using a multi-tone waveform.
Later, in \cite{added_a32}, motivated by the advantages of using multisine waveforms in WET, the author extended the idea to SWIPT systems, with the transmitted symbols resulting from the sum of a deterministic multisine waveform with a modulated multi-carrier waveform, as in orthogonal frequency division multiplexing (OFDM).

%\cite{added_a11} replaced by "Surveys" as [23]
%\cite{added_a25} Will not be cited. Not needed.

The precoders proposed in the present paper are based on statistical CSIT (mean and correlation) and attain power gains very close to the optimal precoder with instantaneous full-CSIT available at the BS. The results are in all setups much superior to the ones obtained by the CSIT-free techniques \cite{IoTJ_2021}. The type of precoding herein proposed provides a linear gain for the system's sum-power as a function of the number of antenna elements at the BS, naturally making the case for the use of massive antenna arrays for WET whenever terminals clustering is possible.
The proposed techniques hold for any waveforms that one may consider, transmitted over flat fading, and therefore the power conversion gains coming from an optimized waveform, such as the multisine waveforms ones in \cite{added_a35,added_Clerckx17} or any of the waveforms listed in \cite[Table 1]{Added_Clerckx20} for different scenarios, will add up to the gains coming from the proposed precoding techniques applied at the BS, preserving the performance gaps found in the current work.

The remainder of the paper starts by setting the system model in section II, then section III lays out the proposed precoding schemes, and does so for increasingly more sophisticated cases. That section also tackles the fairness problem of power allocation among clusters. Section IV presents a baseline reference case that is analytically tractable, and which will serve to validate the numerical simulator used to assess the proposed techniques in section V. Finally, the conclusions are drawn in section VI.

\textit{Notation}: a complex circularly symmetric Gaussian random vector with
mean $\mathbf{m}$ and covariance $\mathbf{R}$ is denoted by $\mathcal{CN} \left(\mathbf{m}, \mathbf{R} \right)$. $\mathbf{A}^H$ denotes the Hermitian transpose of matrix $\mathbf{A}$, $\mathbf{I}_n$ is an identity matrix with $n$ diagonal elements, $\mathbb{1}_M$ is the column vector of $M$ ones, and $\mathbb{E}$ is the expectation operator. $u(x)$ is the Heaviside step function, and $\Gamma(x)$ is the complete Gamma function. The probability density function (PDF) of a random variable $X$ is denoted by $p(x)$ and its power is considered to be equal to its second moment $\mathbb{E}\{X^2\}$, assuming a unit resistor to convert Volt$^2$ to Watt, as it is usual in signal processing. To a set of independent and identically distributed random variables one applies the i.i.d. acronym.

\section{System Model}
\label{Sec:system_model}
The considered scenario for WET is the one in Fig. \ref{Fig:system_setup}, where, without loss of generality regarding the antenna's geometry, a BS equipped with $M$-antennas serves $L$ clusters, each of which encompassing $K$ terminals. The proposals in the paper will be assessed with a uniform linear array (ULA) at the BS, but this option does not preclude the possibility of considering a more general array, such as a uniform planar array (UPA) \cite{added_UPA}, which would give another degree of freedom to the system, allowing for vertical beam steering. Moreover, a scenario considering a different number $K_l$ of devices per cluster is straightforward to assess and can easily be generalized from the model that is here assumed for
illustration purpose. In fact, both precoders that will be derived for multi-cluster systems (in sections \ref{sec: prec_multi-clus} and \ref{sec_numerical_opt}) include the information about the clusters in $K \times M$ matrices that could also be $K_l \times M$ with no impact on how the optimization unfolds.

A Rician MIMO channel \cite[Sec. 5.7]{Hampton}, \cite[Sec. 3.6.1]{HeathLozanoBook}, is considered to exist from the BS to each of the terminals within a cluster.
The baseband model of the signals received by the \textit{K} single-antenna terminals within the \textit{l}-th cluster are aggregated in the vector
\begin{align} \label{Rice_channel}
\mathbf{y}_{l} & \!=\! \Bigg(\underbrace{\sqrt{\frac{\beta(l) \kappa(l)}{1\!+\!\kappa(l)}}}_{\alpha_1(l)}    \mathbf{H}_l^{\text{(LOS)}}\!+ 
\underbrace{\sqrt{\frac{\beta(l)}{1\!+\!\kappa(l)}}}_{\alpha_2(l) } \mathbf{H}_l  \Bigg) \underbrace{\mathbf{P}_l\mathbf{x}_0}_{\mathbf{x}_l} + \mathbf{n}_l,
\end{align}
\noindent where the elements $y_l(k)$ in vector $\mathbf{y}_{l} \in \mathbb{C}^{K \times 1}$ are the received signals at each one of the terminals in the cluster and $\beta(l)$ is the mean large-scale pathloss from the BS to the terminals in cluster $l$.
Matrices $\mathbf{P}_l\in \mathbb{C}^{M \times M}$ will be restricted to be  unitary matrices when deriving the unconstrained optimal beamforming precoder that focus the transmit power onto the $l$-th cluster for efficient WET. That restriction will be dropped when searching for a constrained solution.

The vector onto which the unitary transformation is applied is set to $\mathbf{x}_0 = \sqrt{P_x/M}\mathbf{s}$, where $\mathbf{s} \in \mathbb{C}^{M \times 1}$ is the waveform signal vector in the baseband model, such that $\mathbb{E}\{||s_m||^2\}=1$, for $m=1,\dots, M$, where each $s_m$ represents a waveform, possibly optimized for a particular model of the terminal's EH circuits, such as a deterministic multisine signal \cite{added_a35, added_Clerckx17, added_Clerckx17_lowcomplexity} (a joint optimization of the precoder and of the waveform falls beyond the scope of this work).

As indicated in \eqref{Rice_channel}, the equivalent precoding vector applied to the $M$ antennas is $\mathbf{x}_l \triangleq \mathbf{P}_l\mathbf{x}_0$,
and therefore the transmitted signal to power all the network devices from the BS is $\mathbf{x}_{l} \in \mathbb{C}^{M \times 1}$, with $\mathbb{E}\{\mathbf{x}_0^H\mathbf{x}_0\}$ = $\mathbb{E}\{||\mathbf{x}_0||^2\}$ = $\mathbb{E}\{\mathbf{x}_l^H\mathbf{x}_l\}$ = $\mathbb{E}\{||\mathbf{x}_l||^2\}$ = $P_x$.
$\mathbf{n}_l \sim \mathcal{CN} \left(\mathbf{0}, \sigma_n^2\mathbf{I}_K \right)$, is the vector containing the thermal noises at each single-antenna terminal, each of which with $\sigma_n^2$ power. As usual in the WET literature, this power is considered to be negligible for the purpose of energy harvesting and is neglected in all aspects of the remaining of the paper.

The matrices $\mathbf{H}_l^{\text{(LOS)}}$ and $\mathbf{H}_l$, both $\in \mathbb{C}^{K \times M}$, respectively represent the LOS and the multi-path components of the Rician MIMO channel from the BS to the \textit{l}-th cluster, and $\kappa(l)$ is the Rician factor defining the weights of each component. The clusters' angular width is considered to be narrow and therefore all the terminals approximately experience the same LOS component, sharing the same $\kappa$ and the same slowly varying $\mathbf{H}_l^{\text{(LOS)}}$, which is assumed to be perfectly known at the BS via low-rate feedback.

It is interesting to note that while in the cases in which NOMA is considered to bring some additional capacity to systems using massive MIMO, the users need to be confined to a narrow angular spread \cite{Added_b_NOMA20,Added_b_clerckx2021noma}, which is quite unnatural scenario in cellular communications, the assumption of a narrow angular spread and small range differences seem reasonable in several applications of WET, when particular spots have a higher concentration of EH devices.
Matrix $\mathbf{H}_l$ is a Rayleigh fading matrix with all its elements taken from a circularly symmetric complex Gaussian distribution with all elements $h_{k,m} \sim \mathcal{CN}(0,1)$, while $\mathbf{H}_i^{\text{(LOS)}}$ corresponds to the geometric model \cite[Sec. 5.1]{Hampton}, which typically leads to a matrix of rank one \cite[Sec. 3.6.1]{HeathLozanoBook}. Let $(\mathbf{h}^\text{(LOS)}_{k,l})^T$ be the $k$-th row of $\mathbf{H}_l^{\text{(LOS)}}$, representing the LOS from the BS to the center of the $l$-th cluster.

Because all the terminals in a given cluster approximately experience the same LOS component, $\mathbf{H}_l^{\text{(LOS)}}=\mathbb{1}\cdot(\mathbf{h}^\text{(LOS)}_k)^T$, i.e., all rows replicate the same $(\mathbf{h}^\text{(LOS)}_k)^T = e^{j\varphi_l} [1, e^{j \theta_{1}(\phi_l)} \cdots e^{j \theta_{(M-1)}(\phi_l})]$. The power gain from the $i$-th antenna to the $k$-th terminal is characterized by the same $\beta(l)$ to all users in the $l$-th cluster, given that all antennas of the ULA are co-located, as it has been also considered  in \cite{added_R1, Added_b_Zhang16} and in the results section of \cite{kashyap_feasibility_2016}. $\phi_l$ describes the angular position of the $l$-th cluster, measured in respect to the so-called \textit{endfire} direction of the ULA (in \cite{WCL} it was used the angle measured to the ULA's boresight (also referred to as broadside) \cite[Sec. 3.5]{HeathLozanoBook}). Therefore, $\theta_i(\phi)=2 \pi i d \cos(\phi)/\lambda$, for $\alpha_{i}, i=1, \ldots, M-1$, where $\lambda$ is the wavelength, and $d=\lambda/2$ is the separation between the antenna elements of the ULA. Note that the common phase $\varphi_l$ plays no role in the transferred power to the clusters. A more general case with the $K$ terminals within each cluster randomly distributed within an angular domain $\big[\phi_l-\Delta_\phi ,\ \phi_l+\Delta_\phi \big]$ will also be assessed. In that case, the $k$-th terminal in a given cluster is located at an angle $\phi_k$, taken from a uniform distribution within the angular aperture of $\Delta_\phi$ degrees, and one has different rows $(\mathbf{h}^\text{(LOS)}_k)^T = e^{j\varphi_l} [1, e^{j \theta_{1}(\phi_k)} \cdots e^{j \theta_{(M-1)}(\phi_k)}],$ for each of the $K$ terminals in the $l$-th cluster.

While in most MTC scenarios the line-of-sight component is quasi-static, the fast component $\mathbf{H}_l$ is difficult to be obtained by the BS in a mMTC context, not only because of the power energy limitations of the devices, but also due to the sheer number of terminals. Nonetheless, a central assumption in this paper is that the correlation of the $\mathbf{H}_l$ component is assumed to be known at the BS because it can be slowly updated over time by means of pilots sparsely transmitted by the devices, or via modern machine learning methods \cite{Added_b_ML}. Given the physical proximity of the users within a cluster, according to the geometrical one-ring scattering channel model \cite{adhikary_joint_2013} there is a transmit antenna correlation among the signals received by those $K$ terminals. Thus, the Rayleigh matrix component of the MIMO channel of a cluster has a square transmit correlation matrix $\mathbf{R}_l = \mathbb{E} \{\mathbf{H}_l^{H} \mathbf{H}_l\} \in \mathbb{C}^{M \times M}$ \cite{HeathLozanoBook}, which, via its singular value decomposition (SVD), can be expanded in the form:
\begin{equation} \label{SVD}
\mathbf{R}_l=\mathbb{E}\{\mathbf{H}_l^{H} \mathbf{H}_l\}=\mathbf{U}_l \boldsymbol{\Lambda}_l\mathbf{U}_{l}^H.
\end{equation}

It will be considered that the rank of $\mathbf{R}_l$, $r_l$, is equal to the number of users, similarly to \cite{bana_ultra_2018}. Therefore, $\mathbf{\Lambda}_l$ is only going to have $r_l$ non-zero diagonal elements.
When the channel correlation is known one can take advantage of the KL channel representation, which has been applied in the context of NOMA  \cite{ding_design_2016, Karhunen_Loeve2020}, MIMO spacial division multiplexing \cite{adhikary_joint_2013}, rate-splitting in MU-MIMO systems \cite{dai_clerckx_2015}, and also as an efficient way of generating fading samples with a given correlation matrix \cite{sanguinetti_towards_2019},\cite[Sec. 2.2] {Bjo_Hoy_Sang_2017}. The KL representation allows the channel  from the BS to the users in the $l$-th cluster to be written as
\begin{equation}\label{KL_representation}
\mathbf{H}_l= \mathbf{G}_l \mathbf{\Lambda}_l^{\frac{1}{2}} \mathbf{U}^H_l,
\end{equation}
where $\mathbf{G}_l \in \mathbb{C}^{K \times K}$ is a zero mean and uncorrelated Rayleigh fading matrix, i.e., $\mathbb{E} \{ \mathbf{G}_l \}=\mathbf{0}$, and $\mathbb{E} \{ \mathbf{G}_l^H \mathbf{G}_l \}= \mathbf{I}_K$, the identity matrix with $K$ diagonal elements.
Note that because of the latter property, when plugging \eqref{KL_representation} in \eqref{SVD}, one always gets the same correlation regardless the particular realization of $\mathbf{G}_l$, showing that it only depends on $\boldsymbol{\Lambda}_l$ and $\mathbf{U}_l$ in the KL representation.
$\boldsymbol{\Lambda}_l\in \mathbb{C}^{K \times r_l}$ is a diagonal matrix containing the singular values of $\mathbf{R}_l$ and $\mathbf{U}_{ \textit l} \in \mathbb{C}^{M \times r_l}$ is a matrix containing in its columns the singular vectors of $\mathbf{R}_l$, where $r_k$ is the rank of the correlation matrix. As the diagonal matrix $\mathbf{\Lambda}_l$ can be reduced to a $r_l \times r_l$ matrix, one has $\mathbf{G}_l$ an $K \times r_l$ matrix, and $\mathbf{U}_l$ a $M \times r_l$ matrix.
Hence, the $k$-th user in the $l$-th cluster will experience the channel
\begin{equation}
\mathbf{h}_{k,l}^T= \mathbf{g}_{k,l}^T \mathbf{\Lambda}_l^{\frac{1}{2}} \mathbf{U}_l^H, \quad k=1,\dots,K.
\end{equation}
Note that in the case of uncorrelated fading the $K$ terminals would see a channel $\mathbf{h}_{k,l} \sim \mathcal{CN} \left(0, \mathbf{I}_{M}\right)$.

It is assumed that the BS knows both $\mathbf{H}_l^\text{(LOS)}$ and the autocorrelation in (\ref{SVD}), the latter representing the partial knowledge held about (\ref{KL_representation}). Therefore, while $\mathbf{\Lambda}_l^{\frac{1}{2}}$ and $\mathbf{U}^H_{l}$ can be estimated via the SVD (\ref{SVD}), the fading row vector $\mathbf{g}_{k,l}^T$ is unknown at the BS and cannot be taken in consideration by a precoder. Moreover, because single-antenna terminals are considered, each terminal cannot mitigate this multi-path effect.

The energy harvesting circuits are typically non-linear and the effectively harvested energy will be a fraction of the RF energy available. A rich literature modeling the EH circuits exists, e.g., \cite{added_a31, added_EHcircuits,Added_Clerckx21,added_a35}, and it has been known that different types of RF-DC conversion circuits lead to different optimal waveforms. To assess the proposed precoding techniques, this paper adopts one model that still mimics important features such as the sensitivity and saturation phenomena.
The non-linear EH circuit at the terminals is considered to have a saturation value, $\varpi_2$, and also a minimum power sensitivity, defined by an activation threshold $\varpi_1$, and a conversion factor $\eta$ in its linear region, leading to the following output power, as in \cite{lopez_statistical_2019}:
\begin{equation}\label{conversor}
\Omega(|y_{l,k}|^2)=\left\{
\begin{array}{lrr}
   0, & & |y_{l,k}|^2<\varpi_1 \\
   \eta |y_{l,k}|^2, &  & \varpi_1\le |y_{l,k}|^2<\varpi_2 \\
   \eta\varpi_2,&    & |y_{l,k}|^2\ge \varpi_2
\end{array}
\right..
\end{equation}
\section{Optimal Unitary Precoders for WET}
\label{section: precoders}

A natural goal for WET is the one of maximizing the RF energy available to be harvested by the terminals in each cluster while avoiding energy leakage to other locations. An effective way of achieving that is to apply MIMO precoding at the BS. Using simple analogue phase modulators one can consider signals from a continuous-domain alphabet, with $x_l(m)=\sqrt{P_x}e^{j\theta_m}$, $m=1, \dots, M$, only having to adjust the phases $\theta_{m}$. Given that $\mathbf{P}_l$ is unitary, with $\mathbf{x}_0 = \sqrt{P_x/M}\mathbb{1}_M$, the precoding vector $\mathbf{x}_l =\mathbf{P}_l\mathbf{x}_0$ lives in a $M$-dimensional sphere of norm $\sqrt{P_x}$.

\subsection{Precoding for a single cluster: maximum harvested energy}

The problem of maximizing the mean sum-power, $P_l$, available for EH by all devices in the $l$-th cluster, given $||\mathbf{x}_l||^2=||\mathbf{x}_0||^2=P_x$, can be formalized as:
\begin{argmaxi}
	   {\mathbf{P}_l}{\mathbb{E} \{P_l\}}{}{} \label{opt1}
       \addConstraint{\mathbf{P}_l^H\mathbf{P}_l}{=\mathbf{I}}{}
\end{argmaxi} 
\noindent where the total available energy $\mathbb{E}_l$ at the cluster is the sum of the energies available for harvesting at the $K$ terminals, each of which can collect the power $|y_l(k)|^2$. By using (\ref{Rice_channel}) and (\ref{KL_representation}), the total RF energy available for harvesting by the terminals in cluster $l$, and that should be maximized in (\ref{opt1}), is
\begin{align} \label{opt_expanded}
P_l&= \| \mathbf{y}_l \| ^2 \nonumber\\
&=\lVert \alpha_1(l) \mathbf{H}_l^{\text{(LOS)}}\mathbf{P}_l\mathbf{x}_0 +  \alpha_2(l) \mathbf{G}_l \mathbf{\Lambda}_l^{\frac{1}{2}} \mathbf{U}^H_l \mathbf{P}_l\mathbf{x}_0 \lVert^2
\end{align} 
and, for convenience, let one define:
\begin{equation}
    \mathbf{\Upsilon} \triangleq \alpha_1(l) \mathbf{H}_l^{\text{(LOS)}}\mathbf{P}_l\mathbf{x}_0.
\end{equation}
Consequently, the expectation over the realizations of the Rayleigh fading component $\mathbf{G}_l$ becomes
\begin{align}
  \mathbb{E} \{P_l\} & = \mathbb{E} \{||\mathbf{y}_l||^2\}\nonumber\\
  & = \mathbb{E} \{\lVert \mathbf{\Upsilon}+ \alpha_2(l) \mathbf{G}_l \mathbf{\Lambda}_l^{\frac{1}{2}} \mathbf{U}^H_l \mathbf{P}_l\mathbf{x}_0 \lVert^2 \}\nonumber \\
  & = \mathbb{E} \{ \lVert \mathbf{\Upsilon} \lVert^2 \} + \nonumber\\
  &\qquad\qquad+ \mathbb{E} \{ \lVert \alpha_2(l) \mathbf{G}_l \mathbf{\Lambda}_l^{\frac{1}{2}} \mathbf{U}^H_l \mathbf{P}_l\mathbf{x}_0 \lVert^2 +\nonumber \\
  &\qquad\qquad + 2\alpha_1(l){\mathbf{x}_0}^H {\mathbf{P}_l}^H{(\mathbf{H}_l^{\text{(LOS)}}})^H \times \nonumber \\
  &\qquad\qquad \times \alpha_2(l) \underbrace{\mathbb{E} \{ \mathbf{G}_l \}}_{=\mathbf{0}}  \mathbf{\Lambda}_l^{\frac{1}{2}} \mathbf{U}^H_l \mathbf{P}_l\mathbf{x}_0 \} \nonumber\\
  & = \lVert \mathbf{\Upsilon} \lVert^2 +\nonumber\\
  &\qquad + \alpha_2^2(l) \mathbf{x}_0^H \mathbf{P}_l^H \mathbf{U}_l \mathbf{\Lambda}_l^{\frac{1}{2}} \underbrace {\mathbb{E} \{ \mathbf{G}_l^H \mathbf{G}_l \}}_{=\mathbf{I}} \mathbf{\Lambda}_l^{\frac{1}{2}} {\mathbf{U}}_l^H \mathbf{P}_l\mathbf{x}_0 \nonumber\\
  & = \lVert \mathbf{\Upsilon} \lVert^2 + \lVert \alpha_2(l)\boldsymbol{\Lambda}_l^{\frac{1}{2}}\mathbf{U}^H_l \mathbf{P}_l\mathbf{x}_0 \lVert^2.
\end{align}
Unsurprisingly, this last expression does not depend on $\mathbf{G}_l$, given that the a fast fading unit-power MIMO channel, on average, is a norm-preserving linear transformation. The equivalent optimization problem becomes the one of simultaneously maximizing two squared norms, while considering the power uniformly distributed across the $M$ antennas:
\begin{argmaxi} 
	   {\mathbf{P}_l}{ \lVert \mathbf{\Upsilon} \lVert^2 + \lVert \alpha_2(l) \boldsymbol{\Lambda}_l^{\frac{1}{2}}\mathbf{U}^H_l \mathbf{P}_l\mathbf{x}_0 \lVert^2}{}{} \label{opt1_simplifyed}
	   \addConstraint{\mathbf{P}_l^H\mathbf{P}_l}{=\mathbf{I}}{}
	   %\addConstraint{|x_l(i)|^2}{=\frac{P_x}{M},}{i=1,\ldots,M}
\end{argmaxi} 
$\mathbf{x}_l$ will be used henceforth to indirectly define $\mathbf{P}_l$.

\subsection{Precoding for a single cluster: correlated Rayleigh fading} \label{sec_corr_Rayleigh}

When $\kappa=0$, the optimal unitary precoder for the second term in (\ref{opt1_simplifyed}) can be analytically constructed by splitting the precoding process in two, such that $\mathbf{P}_l=\mathbf{P}^{(1)}_l\mathbf{P}^{(2)}_l$.
By setting $\mathbf{P}^{(1)}_l=\mathbf{U}_l$, the problem simplifies to finding a unit-power vector $\mathbf{P}^{(2)}_l\mathbf{x}_0$ that maximizes $\lVert \boldsymbol{\Lambda}_l^{\frac{1}{2}}\mathbf{P}^{(2)}_l\mathbf{x}_0 \lVert^2$.
Because $\boldsymbol{\Lambda}_l^{\frac{1}{2}}$ is a diagonal matrix with decreasing elements, the maximum singular value, $\sigma_\text{max}$, is always the first element in the diagonal and therefore the maximization requires $\mathbf{P}^{(2)}_l\mathbf{x}_0 =[1, 0 \dots 0]^T$.
In short, $\mathbf{x}_l=\mathbf{P}^{(1)}_l\mathbf{P}^{(2)}_l\mathbf{x}_0=\mathbf{U}_l [1, 0 \dots 0]^T$, concluding that when there is no LOS and only the multi-path (MP) channel component exists, the optimal transmission is defined by the first column of $\mathbf{U}_l$, obtained in (\ref{SVD}):
\begin{equation} \label{opt_MP}
\mathbf{x}_l^{*\text{(MP)}}= \sqrt{P_x} \mathbf{U}_l(:,1),
\end{equation}
\noindent using MATLAB\textsuperscript{\textregistered} notation. Implicitly, $\mathbf{P}^{(2)}_l$ applies a unitary rotation to $\mathbf{x}_0$ to generate $[1,0 \dots 0]^T$, such that $\mathbf{P}_l^{(2)}$ is a rotation matrix, and therefore the total precoding matrix $\mathbf{P}_l=\mathbf{P}^{(1)}_l\mathbf{P}^{(2)}_l$ is unitary.

\subsection{Precoding for a single cluster: correlated Rician fading} \label{sec_corr_Rice}

In practical WET setups, the powering BS is usually located close to the clusters of users and therefore having a strong LOS component with $\kappa>1$ is a likely situation.
When both terms in (\ref{opt1_simplifyed}) contribute to $\mathbb{E} \{P_l \} = \mathbb{E} \{||\mathbf{y}_l||^2\}$, the optimum $\mathbf{x}_l^*$ is obtained by maximizing all the coordinates $y_l(k)$, which, according to \eqref{opt1_simplifyed}, corresponds to maximizing $K$ inner products of $\mathbf{x}_l$ with the $k$-th row of $\mathbf{H}_l^\text{LOS}$, denoted by $(\mathbf{h}_k^{\text{(LOS)}})^T$ and also other $r_l$ inner products in the norm of the second term of \eqref{opt1_simplifyed}.
Defining 
\begin{equation}
\mathbf{M}_l \triangleq \boldsymbol{\Lambda}_l^{\frac{1}{2}}\mathbf{U}^H_l \in \mathbb{C}^{r_l, \times r_l},
\end{equation}
\noindent and $\mathbf{m}_{k}^T$ being its $k$-th row, the total power collected by the cluster is (for simplicity the $l$ index is dropped in the vectors)
%\begin{equation}
%\begin{aligned}
%    \mathbb{E} \{||\mathbf{y}_l||^2\} & =  \alpha_1^2(l)\big\lVert %\mathbf{H}_l^{\text{(LOS)}}\mathbf{P}_l\mathbf{x}_0 \lVert^2 + \alpha_2^2(l) \lVert %\mathbf{M}_l \mathbf{P}_l\mathbf{x}_0 \big\lVert^2 \\
%    & = \alpha_1^2(l) \Big( |(\mathbf{h}_1^{\text{(LOS)}})^T %\mathbf{P}_l\mathbf{x}_0|^2 + \dots + |(\mathbf{h}_K^{\text{(LOS)}})^T %\mathbf{P}_l\mathbf{x}_0|^2\Big) + \\
%    & + \alpha_2^2(l) \Big( |(\mathbf{m}_1^T \mathbf{P}_l\mathbf{x}_0|^2 + \dots + %|(\mathbf{m}_{r_{l}}^T \mathbf{P}_l\mathbf{x}_0|^2  \Big),
%\end{aligned}
%\end{equation}
\begin{align} \label{opt_Rice}
    P_l & =  \alpha_1^2(l)\big\lVert \mathbf{H}_l^{\text{(LOS)}}\mathbf{P}_l\mathbf{x}_0 \lVert^2 +
    \alpha_2^2(l) \lVert \mathbf{M}_l \mathbf{P}_l\mathbf{x}_0 \big\lVert^2\nonumber \\
    & = \alpha_1^2(l) \sum_{k=1}^{K} \big| (\mathbf{h}_k^{\text{(LOS)}})^T \mathbf{P}_l\mathbf{x}_0 \big|^2 +\nonumber \\
    &\qquad\qquad + \alpha_2^2(l) \sum_{k=1}^{r_l} \big| \mathbf{m}_k^T \mathbf{P}_l\mathbf{x}_0 \big|^2,
\end{align}
which can be written in the form
%\begin{equation}
\begin{align} \label{opt_Rice_simplifyed}
    P_l & = \Big\|
    \begin{bmatrix}
    \alpha_1(l)& 0\\
    0&\alpha_2(l)
    \end{bmatrix}
    \begin{bmatrix}
    \mathbf{H}_l^\text{(LOS)}\\
    \mathbf{M}_l
    \end{bmatrix} \mathbf{P}_l\mathbf{x}_0 \Big\|^2\nonumber \\
    & = \Big\|  \underbrace{\begin{bmatrix}
    \alpha_1(l) \mathbf{H}_l^\text{(LOS)}\\
    \alpha_2(l) \mathbf{M}_l
    \end{bmatrix}}_{\mathbf{A}_l} \mathbf{P}_l\mathbf{x}_0  \Big\|^2.
%\end{aligned}
\end{align}

Both the LOS and the multi-path components contribute to the problem, such that $\mathbf{A}_l \in \mathbb{C}^{2K \times M}$ is the maximum dimension of this matrix (depending on the rank of $\mathbf{H}_l^\text{(LOS)}$ and the rank of $\boldsymbol{\Lambda}_l$, $r_l$). From (\ref{opt_Rice_simplifyed}), one can conclude that the maximum possible harvested energy corresponds to the largest singular vector of $\mathbf{A}_l$, $\sigma_{\text{max}}$, and is obtained when the transmitted signal, $\mathbf{x}_l$, is the (unit power) singular vector $\mathbf{u}_{\text{max}}^{(\mathbf{A}_l)}$ associated to $\sigma_\text{max}$. When taking in consideration both the LOS channel and the second order statistics of the multi-path channel, the optimal solution to (\ref{opt1_simplifyed}) is therefore
\begin{equation} \label{xl_optimal}
    \mathbf{x}_l^*=\sqrt{P_x}\mathbf{u}_{\text{max}}^{(\mathbf{A}_l)},
\end{equation}
which is a vector living on the $M$-sphere of norm $\sqrt{P_x}$, and $\mathbf{P}^{(2)}_l$ is implicitly a rotation matrix acting on $\mathbf{x}_0$, as in the previous subsection.

\subsection{Precoding for multiple clusters} \label{sec: prec_multi-clus}
The maximization of the average energy available to be harvested by the system's devices in a multi-cluster system, $\mathbb{E}\{P\}$, can be attained by constructing the stacked vector $\mathbf{y}=[\mathbf{y}_1 \dots \mathbf{y}_l \dots \mathbf{y}_{L}]^T$ with all the signals at all the terminals in all the clusters. The problem then becomes the one of maximizing the (squared) norm of this $KL$-dimensional vector. With a similar manipulation to the one in \eqref{opt_Rice}, one can obtain the optimal precoder for the multi-cluster case ($L\geq1$), which is the same of finding the optimal precoding vector $\mathbf{x}^* = \mathbf{P}^*\mathbf{x}_0$. Hence, 
\begin{equation} \label{opt_multi_clusters}
    P= \lVert \mathbf{y} \lVert^2 = \left\|
    \underbrace{\begin{bmatrix}
    \alpha_1(1) \mathbf{H}_1^\text{(LOS)}\\
    \alpha_2(1) \mathbf{M}_1\\
    \vdots\\
    \alpha_1(l) \mathbf{H}_l^\text{(LOS)}\\
    \alpha_2(l) \mathbf{M}_l\\
    \vdots\\
    \alpha_1(L) \mathbf{H}_L^\text{(LOS)}\\
    \alpha_2(L) \mathbf{M}_L\\
    \end{bmatrix}}_{\mathbf{A}} \underbrace{\mathbf{P}\mathbf{x}_0}_{\mathbf{x}}  \right\|^2,
\end{equation}
is maximized when transmitting the right-singular vector associated to the maximum singular value of the SVD of the concatenated matrix $\mathbf{A} \in \mathbb{C}^{2KL \times M}$ (assuming that both $\mathbf{H}_l^\text{(LOS)}$ and $\mathbf{\Lambda}_l$ are full-rank), which is also a  vector living on the $M$-sphere of norm $\sqrt{P_x}$:
\begin{equation}
    \mathbf{x}^*=\sqrt{P_x}\mathbf{u}_{\text{max}}^{(\mathbf{A})}\label{x_optimal}.
\end{equation}

The solution (\ref{x_optimal}) takes into account the whole multi-cluster system, and for that reason it depends on all the correlation matrices $\mathbf{R}_l$ and on all the LOS channels $\mathbf{H}_l^\text{(LOS)}$.

The complexity of the proposed precoding scheme is chiefly determined by the SVD of the autocorrelation and the SVD of $\mathbf{A}_l \in \mathbb{C}^{2KL, \times M}$, which has a time complexity $\mathcal{O}\left(4KLM^{2}+M^{3}+M+2KL M\right)$ and a storage complexity $\mathcal{O}\left(3M^{2}+3M+4KLM\right)$ by using the so-called truncated SVD algorithm \cite{svd_complex_2019}.

\textit{Remark}: \eqref{opt_multi_clusters} is applicable to a MU-MISO setup in which device clustering is possible, which implies a MU-MIMO setup with more than one EH terminal. When clustering is not possible in a MU-MIMO system \cite{Thudugalage.2016}, or in in the case of a point-to-point single-user (SU) MIMO \cite{Added_b_Zhang16,added_R1}, the solution to the problem of maximizing the available RF power is also a SVD problem involving the corresponding Wishart matrix. In the case of the Rician model in \eqref{Rice_channel}, the distribution of the maximum eigenvector is known \cite{Added_b_Palomar}, as well for the case of a Rayleigh MIMO channel \cite{Added_b_Eigen_Rayleigh}. Moreover, in the simpler case of SU-MISO, the matrix problem collapses to the well-known vector problem of coherent combining \cite{kashyap_feasibility_2016}.

\subsection{Constrained precoding for efficient non-linear EH and inter-cluster fairness} \label{sec_numerical_opt}

Although the energy available for a cluster is $\mathbb{E} \{||\mathbf{y}_l||^2\} $, the sensitivity and saturation thresholds of the non-linear transfer function of the harvesting circuit, described in (\ref{conversor}), induce energy waste unless $\varpi_1^2\leq\mathbb{E} \{|y_l(k)|^2\}\leq\varpi_2^2$, at each terminal in the $l$-th cluster. In order to minimize that, the precoder should also take into account those additional constraints. An additional affect of adding the lower bound constraint for each of the $KL$ terminals is the induction of fairness among the partial sum-power of each cluster. This is a natural consequence of the fact that the optimization is blind to which cluster each terminal belongs to. The constraint optimization problem is set in the following, where $\mathbf{P}$ is no longer required to be unitary because a solution for $\mathbf{P}$ spat out by the numerical solver may observe the set of constraints without having to reach the maximum allowed transmit power. One can simply state the problem as:
\begin{argmaxi} 
	   {\mathbf{x}}{
	   \begin{aligned}
	   &\sum_{l=1}^L \Big( \lVert \alpha_1(l) \mathbf{H}_l^{\text{(LOS)}}\mathbf{x}\lVert^2 + \lVert \alpha_2(l) \mathbf{M}_l\mathbf{x}\lVert^2 \Big)
	   \end{aligned}}{}{} \label{opt2}
	   \addConstraint{||\mathbf{x}||^2}{\leq P_x}
	   \addConstraint{\varpi_1\leq}{\mathbb{E}\{|y_l(k)|^2\}\leq}{\varpi_2,}{\quad k=1,\ldots,K}
	   \addConstraint{}{}{\text{and},}{\quad l=1,\ldots,L.}
\end{argmaxi} 
This is a non-linear programming problem with non-linear inequality constraints on each component which can be numerically solved by converting the problem from the complex domain to the real domain. To that end, the MIMO real-equivalent model, well-known context of MIMO detection \cite{Thesis}, is applied to the objective function in (\ref{opt2}), so that a real-valued solution is found by casting the problem in a similar manner to (\ref{opt_multi_clusters}) and then applying the real-equivalent model to
\begin{equation}
\mathbf{A}\mathbf{x} \Leftrightarrow
\left[\begin{array}{cc}
\Re\left(\mathbf{A}\right) & -\Im\left(\mathbf{A}\right) \\
\Im\left(\mathbf{A}\right) & \Re\left(\mathbf{A}\right)
\end{array}\right]\left[\begin{array}{c}
\Re\left(\mathbf{x}\right) \\
\Im\left(\mathbf{x}\right)
\end{array}\right].
\end{equation}
In the end, the solution is converted back to a $\mathbf{x}\in\mathbb{C}^M$.

The constraint added to the problem in \eqref{opt2} requires setting an expression for the \textit{average} energy at each terminal since its \textit{instantaneous} value cannot be estimated by the BS given that it does not have access to the instantaneous realizations of $\mathbf{G}$.
The constraints on the average energy received by each terminal, $\mathbb{E} \{|y_l(k)|^2\}$ can only take into account expectations in the context of the KL decomposition channel model. Using (\ref{Rice_channel}) and (\ref{KL_representation}), while $\mathbf{g}_k^T$ is the $k$-th \textit{row} of $\mathbf{G}$, and dropping the cluster index in the \textit{rows} of $\mathbf{H}^{\text{(LOS)}}$, the per-terminal average energy, when given some precoding matrix $\mathbf{P}$, is
\begin{align} \label{components_power}
    \mathbb{E} \{|y_l(k)&|^2\ | \mathbf{P}\} \nonumber \\
    & = \mathbb{E} \Big\{\alpha_1^2(l) \big|(\mathbf{h}_k^{\text{(LOS)}})^T \mathbf{P}\mathbf{x}_0\big|^2\Big\}\nonumber\\
    & \qquad\qquad + \mathbb{E} \Big\{\alpha_2^2(l)\big|\mathbf{g}_k^T \boldsymbol{\Lambda}_l^{\frac{1}{2}}\mathbf{U}^H_l \mathbf{P}\mathbf{x}_0\big|^2 \Big\} \nonumber\\
    & =\alpha_1^2(l) \big|(\mathbf{h}_k^{\text{(LOS)}})^T \mathbf{x}\big|^2 + \alpha_2^2(l) \mathbb{E} \Big\{\big|\mathbf{g}_k^T \mathbf{M}_l \mathbf{x}\big|^2 \Big\} \nonumber\\
    & =\alpha_1^2(l) \big|(\mathbf{h}_k^{\text{(LOS)}})^T \mathbf{x}\big|^2 +  \alpha_2^2(l) \mathbb{E} \Big\{\big|\mathbf{g}_k^T \mathbf{v}_l\big|^2\Big\}\nonumber \\
    & =\alpha_1^2(l) \big|(\mathbf{h}_K^{\text{(LOS)}})^T \mathbf{x}\big|^2 + \alpha_2^2(l) \mathrm{var} \big\{\mathbf{g}_k^T \mathbf{v}_l\big\}\nonumber\\
    & =\alpha_1^2(l) \big|(\mathbf{h}_k^{\text{(LOS)}})^T \mathbf{x}\big|^2 + \alpha_2^2(l) \mathbf{v}_l^T \underbrace{\mathrm{cov}\{\mathbf{g}_k\}}_{=\mathbf{I}}\mathbf{v}_l\nonumber\\
    & =\alpha_1^2(l) \big|(\mathbf{h}_k^{\text{(LOS)}})^T \mathbf{x}\big|^2 + \alpha_2^2(l) \big\|\mathbf{M}_l \mathbf{x}\big\|^2,
\end{align}
where $\mathbf{v}_l\triangleq\mathbf{M}^H_l\mathbf{P}\mathbf{x}_0=\mathbf{M}^H_l\mathbf{x}$, which is a factor only depending on the cluster's correlation $\mathbf{R}_l$, while the fast fading $\mathbf{g}_i$ plays no role in determining the average power. In \eqref{components_power}, it was also used the fact that the variance of an inner product between a random vector $\mathbf{a}$ and a deterministic vector $\mathbf{b}$ observes $\mathrm{var}(\mathbf{a}^T \mathbf{b})=\mathbf{b}^T \mathrm{var}(\mathbf{a})\mathbf{b}$.

In the case when full CSIT exists (which will be considered for benchmarking in the next section), because the instantaneous value of $\mathbf{G}$ is known, the exact $\mathbf{y}_l$ is also known, as defined in \eqref{Rice_channel}, the expectation $\mathbb{E}\{|y_l(k)|^2\|$ in (\ref{opt2}) can be replaced by the exact received power at each terminal so that the second constraint becomes $\varpi_1\leq |y_l(k)|^2\leq\varpi_2$, for $k=1,\ldots,K$, and $ l=1,\ldots,L$.

\section{The baseline reference case}\label{Sec: baseline}

\newtheorem{prop}{Proposition}

The previous section presented derived the beamforming vectors $\mathbf{x}_l$, including the two numerical precoders that solve the energy-constrained problem (with partial- and full-CSIT). In the follow up section \ref{sec_numerical_res} the performance of these precoders will be compared with the one attained by the AA and SA CSIT-free schemes. While the latter schemes can have an analytical interpretation \cite{lopez_statistical_2019}, the former are analytically intractable. However, there is a particular situation that has an analytical interpretation: the one with a pure Rayleigh channel, i.e., without LOS ($\kappa=0$), and where the fading  is uncorrelated. This situation is considered as the baseline reference case: the i.i.d. Rayleigh fading channel is represented by a MIMO matrix with circularly symmetric complex Gaussian entries with zero mean and unit variance. In this situation $\mathbf{h}_{t,l} \sim \mathcal{CN} \left(0, \mathbf{I}_M \right)$, with both real and imaginary components of each element $h_{t,l}(m)$ having each a Gaussian distribution $\mathcal{N} \left(0, \frac{1}{2}\right)$, such that $h_{t,l}(m) \sim \mathcal{CN} \left(0, 1 \right)$, for $m=1,\dots,M$. This setup has an analytical interpretation chiefly based on the properties of the Gamma distribution \cite{Heath2011,DistMaMIMO_2018}.

\vspace{0.2cm}
\textbf{Definition 1}: A Gamma random variable with finite \textit{shape} parameter $k>0$ and finite \textit{scale} parameter $\theta>0,$ denoted as $\Gamma(k, \theta)$, has PDF
\begin{equation}
	p(x,k,\theta)=x^{k-1} \frac{e^{-x / \theta}}{\theta^{k} \Gamma(k)}u(x), 
\end{equation}
with mean $k\theta$ and variance $k \theta^{2}$.

Hence, if the received power is described by a Gamma distribution, for a linear increase of the mean value there will exist a larger spread of the power domain (which can be also linear or become quadratic, depending whether it is the shape or the scale parameter that is changing).

The link budget is given by the equation $P_r = P_x + G_a + G_P - \beta$ (where the gain of the ULA's antenna elements is considered to be $G_a$ dB, as in \ref{sec_numerical_res}), and therefore one can define an \textit{equivalent power gain}
\begin{equation}	
	\beta_{eq}=\frac{\beta}{P_xG_a},
\end{equation}
\noindent and herein consider a system with unit total transmit power ($P_x=1$) and an ULA with antenna elements with no gain.
	
\begin{prop}
	If $H_m$ are i.i.d random variables with a Gaussian PDF, with $H_m \sim \mathcal{CN} \left(0, 1 \right)$, then the sum $Y=\sum_{m=1}^M H_m$ is also Gaussian distributed with $Y_m \sim \mathcal{CN} \left(0, M \right)$.
\end{prop}

Thus, given that under Rayleigh fading each element of $\mathbf{H}$ (where the cluster index is dropped in a non-LOS situation) is $h_{t}(m) \sim \mathcal{CN} \left(0, 1 \right)$, and reminding that $\mathbf{x}_0 = \sqrt{\frac{1}{M}} \mathbb{1}_M$, one has $\mathbf{h}_t^T \mathbb{1}_M\ \sim \mathcal{CN}\left(0, M \right)$, and the signal at each terminal is $y_t =\frac{1}{M}\mathbf{h}_t^T\mathbb{1}_M$, with variance $\frac{1}{M}M=1$, and therefore its distribution is $y_t \sim \mathcal{CN}\left(0, 1 \right)$, for $t=0, \dots, K$.

\begin{prop}
	If $Y$ is a random variable with $Y \sim \mathcal{CN} \left(0, 1\right)$, then $|Y|$ is Rayleigh distributed, i.e., $p(y,\sigma_R)=\frac{y}{\sigma_R^2}e^{-y^2/(2\sigma_R^2)}$, with parameter $\sigma_R^2=1/2$, and its second moment (or power) is $\mathbb{E}\{Y^2\}=2\sigma_R^2\Gamma(2)=1$ \cite[Sec. 2.B.2]{Simon_2006}, also using the fact that $\Gamma(2)=1$. Moreover, $X=\lvert Y \rvert^2$ has an exponential distribution $p(x,\lambda)=\lambda e^{-\lambda x} u(x)$ with rate parameter $\lambda=1$, holding an average $\mathbb{E}\{X\}=\mathbb{E}\{\lvert Y \rvert^2\}=\lambda^{-1}=1$, as seen before. By virtue of Definition 1, this PDF can also be written as $|Y|^2 \sim  \Gamma(k=1,\theta=1)$.
\end{prop}

Consequently, the RF power available at each terminal is $\mathbb{E}\{\lvert y_t \rvert^2\}=\mathbb{E}\{\frac{1}{M}\lvert\mathbf{h}_t^T\mathbb{1}_M\rvert^2\}$, which is exponentially distributed with $\lambda=1$, at all terminals $t=0, \dots, K$, or equivalently, $\mathbb{E}\{\lvert y_t\rvert^2\} \sim \Gamma(1,1)$.

\begin{prop} \label{sum_of_exps}
	If $Y_t$ are i.i.d. with  PDF $p(y,\lambda)=\lambda e^{-\lambda y}$, then $Z=\sum_{m=1}^K Y_t$ is Gamma-distributed with $Z \sim \Gamma(k=K,\theta=\lambda)$.
\end{prop}

Hence, when experiencing an i.i.d. Rayleigh channel, the partial sum-power available for the $K$ terminals within the $l$-th cluster, is $\lVert \mathbf{y}_{l} \lVert^2  \sim \Gamma\left(k=K, \theta= 1 \right)$.

\begin{prop} \label{scaling_gamma}
	If a random variable $X$ holds a Gamma distribution $\Gamma(k,\theta)$, then, for a scalar $a>0$, $aX$ has a distribution $\Gamma(k, a \theta)$.
\end{prop}

For this reason, a link to a cluster having an equivalent power gain $\beta_{eq}$, the partial sum-power available to the $K$ terminals is distributed according to $\Gamma\left(K, \beta_{eq} \right)$

In order to consider the general situation where different clusters have different $\beta_l$, one will need one more tool given by the so-called \textit{second order Gamma approximation} of the sum of gamma functions \cite{Heath2011} to be given in proposition 6. Note that if all $beta_l$ are the same, then by reapplying proposition 3, one could immediately state $\lVert \mathbf{y}_{l} \lVert^2  \sim \Gamma\left(KL, \beta_{eq} \right)$, given that all users could be considered in the same cluster.

\begin{prop} \cite{Heath2011} \label{2nd_order_approx_a}
	Suppose that $\left\{Y_{t}\right\}$ are independent $\Gamma\left(k_{t}, \theta_{t}\right)$ random variables. The sum $\mathcal{Y}=\sum_{t} Y_{t}$ has mean, variance, and second moment, respectively given by
\end{prop}

\begin{equation}
	\begin{aligned}
		&\mathbb{E} \{\mathcal{Y} \}=\sum_{l} k_{l} \theta_{l},\\
		&\operatorname{var} \{ \mathcal{Y} \}=\sum_{l} \operatorname{var}\left(Y_{l}\right)=\sum_{t} k_{l} \theta_{l}^{2},\\ 
		&\mathbb{E} \{ \mathcal{Y}^2 \}=\sum_{l} k_{l} \theta_{l}^{2}+\Big(\sum_{l} k_{l} \theta_{l}\Big)^{2}.
	\end{aligned}
\end{equation}

\begin{prop} \cite{DistMaMIMO_2018,Heath2011} \label{2nd_order_approx_b}
	If $Y_l$ are independent random variables, each one having a Gamma distribution $\Gamma(k_l,\theta_l)$, their sum $Y=\sum_{l=1}^L Y_l$ can be approximated by a Gamma distribution having its shape and scale parameters given by
\begin{equation}
	k_{y}=\frac{\left(\sum_{l=1}^L k_{l} \theta_{l}\right)^{2}}{\sum_{l=1}^L k_{l} \theta_{l}^{2}} \quad \text { and } \quad \theta_{y}=\frac{\sum_{l=1}^L k_{l} \theta_{l}^{2}}{\sum_{l=1}^L k_{t} \theta_{l}} \label{sum_gammas_1},
\end{equation}
which is known as the \textit{second order Gamma approximation} of the sum that builds up $\mathcal{Y}$ \cite{Heath2011}.
\end{prop}

In the case where all the channels to the terminals in the \textit{l}-th cluster are identically distributed, with $\mathbf{h}^T_{t,l} \sim \mathcal{CN} \left(0, \beta_l \mathbf{I}_M \right)$, then the partial sum-powers at each cluster are Gamma-distributed $\Gamma \left(k_l=K,\theta_l=\beta_{eq} \right)$, for $t=l,\dots,L$, and \eqref{sum_gammas_1} particularizes to
\begin{equation} \label{sum_clusters}
	k_{y}=\frac{\left(\sum_{l=1}^L K \beta_{eq}\right)^{2}}{\sum_{l=1}^L K \beta_{eq}^2}=KL, \hspace{0.3em} \text {and} \hspace{0.3em} \theta_{y}=\frac{\sum_{l=1}^L K \beta_{eq}^2}{\sum_{l=1}^L K \beta_{eq}}=\beta_{eq},
\end{equation}
leading to the following PDF for the system's sum-power:
\begin{equation} \label{eq: AA_gamma_sum-power}
	p(P)=\Gamma \left(KL, \hspace{0.3em} \beta_{eq} \right).
\end{equation}

\begin{prop} \cite{DistMaMIMO_2018}
	For a $K$-dimensional vector with $\lVert \mathbf{y} \lVert^2  \sim \Gamma\left(k_y, \theta_y\right)$, its projection onto $c$ coordinates holds a power distribution $\Gamma(\frac{c}{K}k_y,\theta_y)$, meaning that its mean is $\frac{c}{K}k\theta$ and its variance is $\frac{c}{K}k_y\theta_y^2$, and therefore has a power given by its second moment: $P = \frac{c}{K}k_y\theta_y^2+\left(\frac{c}{K}k_y\theta_y\right)^2$, which amounts to the power available to a subset of $c$ terminals, when $k_y$ and $\theta_y$ are the ones in \eqref{eq: AA_gamma_sum-power}.
\end{prop}
The last proposition establishes the partial sum-power available for EH by a subset of devices in a given cluster.

\subsection{SA scheme}

The second order Gamma approximation method can also be used to find the expression for the PDF of the system's sum-power when using the SA technique, providing a simpler proof than the one given in \cite{IoTJ_2021}.

Under the SA scheme, the received signal at the $t$-th terminal results from a time-slot based accumulation of power, such that $\mathbb{E}\{ \lvert y_t \rvert^2\} = \frac{\beta_{eq}}{M}\sum_m^M \mathbb{E} \{ \lvert h_{t,l}(m)\rvert )^2 \}$, for $t=1, \dots, K$. From proposition \ref{sum_of_exps}, $\sum_m^M \mathbb{E} \{ \lvert h_{t,l}(m)\rvert )^2 \} \sim  \Gamma \left( M, 1 \right)$, and from proposition \ref{scaling_gamma} comes that $\mathbb{E} \{ \lvert y_t \rvert^2\} = \Gamma \left( M , \frac{\beta_{eq}}{M} \right)$.
Within a cluster, by applying proposition \ref{2nd_order_approx_b} with $k_y=M$ and $\theta_y=\frac{\beta_{eq}}{M}$, the cluster's partial sum-power has distribution $\Gamma \left(MK, \frac{\beta_{eq}}{M} \right)$.
Finally, for the whole system, by reapplying proposition \ref{2nd_order_approx_b} with $k_y=MK$ and $\theta_y=\frac{\beta_{eq}}{M}$, the system's sum-power with SA has the following PDF:
\begin{equation}\label{eq: SA_gamma_sum-power}
	p(P)=\Gamma \left(MKL, \hspace{0.3em} \frac{\beta_{eq}}{M} \right). 
\end{equation}
It should be noted that by using SA, differently from \eqref{eq: AA_gamma_sum-power}, the shape parameter in \eqref{eq: SA_gamma_sum-power} grows with $M$, linearly increasing the mean power, while the scale parameter is reduced by $M$, reducing the variance quadratically (as per definition 1 and proposition \ref{2nd_order_approx_a}), which explains why the SA scheme is preferable to AA when $\kappa=0$.

\section{Numerical Results}\label{sec_numerical_res}
In this section, one numerically assesses the proposed precoding (or beamforming) techniques based on the KL channel decomposition with limited CSIT, namely the correlation of the channels from the BS to each of the $L$ clusters. Both the unconstrained optimal analytical precoding for multiple clusters, given by \eqref{x_optimal}, and the numerically obtained constrained precoder given by \eqref{opt2}, will be assessed and compared with the case when full-CSIT is available. The precoder when full-CSIT is available is the one that maximizes 
\begin{equation} \label{opt_full-csi}
    P= \lVert \mathbf{y} \lVert^2 = \left\|
    \underbrace{\begin{bmatrix}
    \alpha_1(1) \mathbf{H}_1^\text{(LOS)}\\
    \alpha_2(1) \mathbf{H}_1\\
    \vdots\\
    \alpha_1(l) \mathbf{H}_l^\text{(LOS)}\\
    \alpha_2(l) \mathbf{H}_l\\
    \vdots\\
    \alpha_1(L) \mathbf{H}_L^\text{(LOS)}\\
    \alpha_2(L) \mathbf{H}_L\\
    \end{bmatrix}}_{\mathbf{B}} \mathbf{P}\mathbf{x}_0  \right\|^2,
\end{equation}
which is given by the singular value associated to the largest singular value of $\mathbf{B}$,
\begin{equation}
    \mathbf{x}^*=\sqrt{P_x}\mathbf{u}_{\text{max}}^{(\mathbf{B})}\label{x_full-csi_optimal}.
\end{equation}

The primal method to to assess the proposed techniques is to focus on the PDF of the sum-power available for harvesting at the terminals, $p(P)$. Subsequently the analysis focuses on how the Rician factor, the dimension of the antenna array at the BS, the angular position of a cluster, and the number of clusters impact on the energy available for harvesting by the terminals when using the proposed optimal precoder defined in \eqref{xl_optimal}, for the single-cluster case, and in  \eqref{x_optimal} for the multi-cluster case.

\subsection{Sum-power distributions: setup}

\begin{figure}[t]
\centering 
\includegraphics[width=1\columnwidth, clip=true, draft=false]{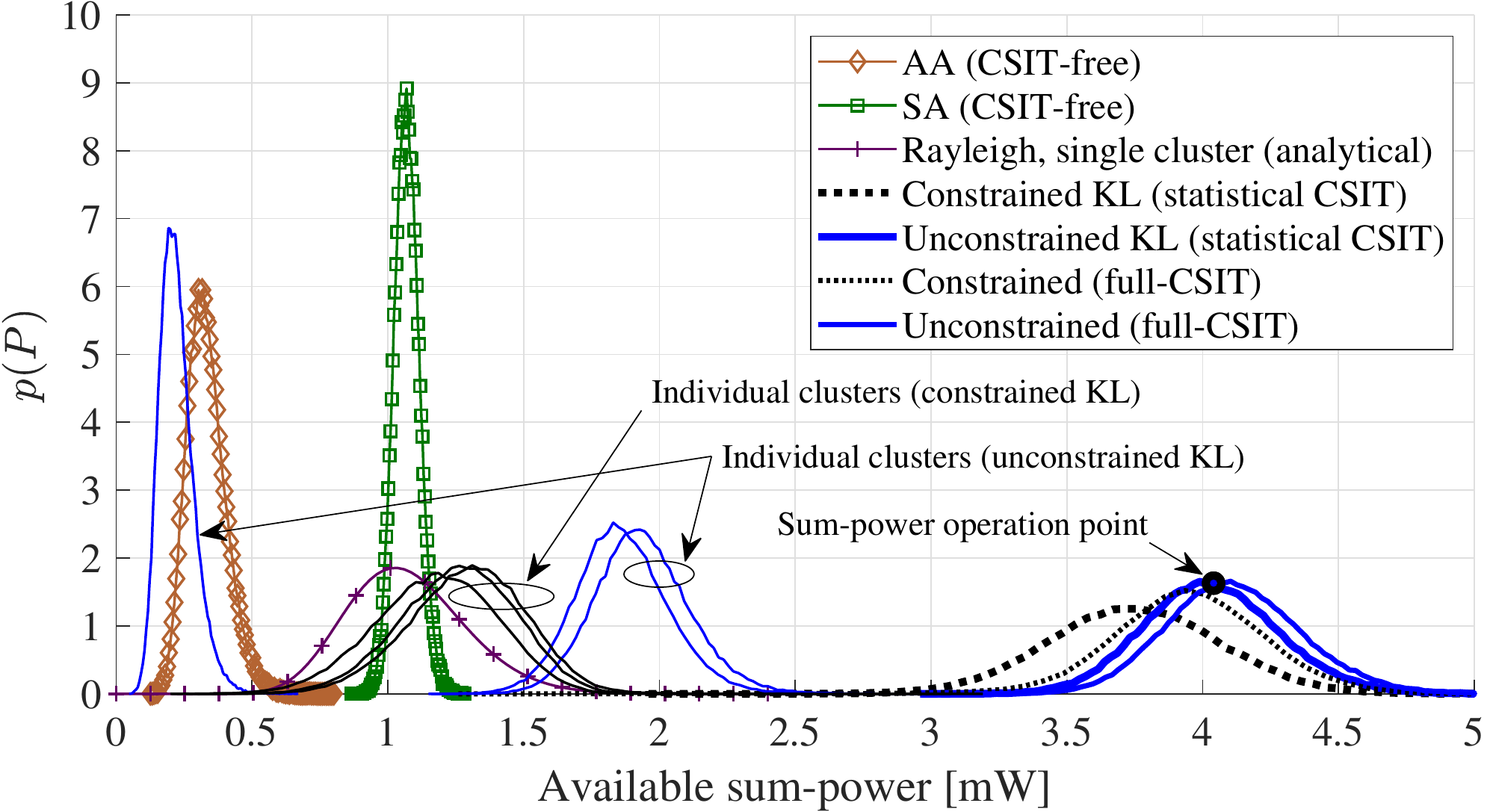}
\caption{Sum-power distribution for different precoding schemes, and partial power distribution per cluster ($M=8$ antennas, $K=8$ users per cluster, and $\kappa=5$) with $L=3$ clusters located at $\phi=\{0, 30, 70\} \text{ degrees}$. Includes the (analytical) reference case without clustering and with uncorrelated Rayleigh fading without LOS.
} \label{fig:distributions}
\end{figure}
The first configuration to be assessed involved one single cluster only, initially with correlated Rayleigh fading only, assessing the precoder devised in section \ref{sec_corr_Rayleigh}, and then adding a LOS component, assessing the precoder devised in section \ref{sec_corr_Rice}. These initial results are not shown due lack of space. The first results here presented consider a multi-cluster system. The observations with one cluster were in line with the ones presented for a multi-cluster system in terms of the impact of the system's parameters in the overall PDF of the sum-power. The set of distributions presented in Fig. \ref{fig:distributions} were obtained for a system with $L=3$ clusters with $K=8$ users per cluster and a BS with $M=8$ antennas (hence, a WET system with a total of 24 terminals), which is still far from a massive MIMO setup but is a typical MIMO array. Without loss of generally, the mean pathloss to each cluster, $\beta$, was considered equal to all clusters, and a LOS characterized by $\kappa=5$ was also considered equal in all clusters. Such Rician factor is adequate to a mMTC system as the one in Fig. \ref{Fig:system_setup}. In \cite{WCL} a $\kappa=10$ was considered, accounting for systems with a quite large LOS component, while in \cite{lopez_statistical_2019} a $\kappa=3$ was rather used.
\begin{table}[t]
    \caption{System's parameters for the operation point in Fig. \ref{fig:distributions}}
    \begin{center}
        \begin{tabular}{ |c| c| }
        \hline
        Num. clusters, $L$ & 3 \\  
        \hline
        Users/cluster, $K$ & 8 \\ 
        \hline
        Antennas BS, $M$ & 8 \\  
        \hline
        Rician factor. $\kappa$ & 5    \\
         \hline
        Tx. power, $P$ & 10 W  \\
         \hline
        Tx. antenna gain & 10 dB \\
         \hline
        Pathloss (channel gain), $\beta$ & $- 63.5$ dB \\
         \hline
        Optimal precoder gain, $G_P$ & $19.6$ dB \\
         \hline
        EH efficiency, $\eta$ & $0.25$\\
        \hline
        \end{tabular}
    \end{center}
    \label{tab:1_parameters}
\end{table}
Fig. \ref{fig:distributions} shows the PDFs of the sum-powers obtained by all the techniques and the analytical curve for the baseline reference case is also added, as defined by \eqref{eq: AA_gamma_sum-power} , with shape parameter $KL=24$ and scale parameter $\beta=10^{-\frac{63.5}{10}}$.

Considering the experimental findings in \cite{Mayaram2008}, the present paper considers $\varpi_1=-22$ dBm ($6.30\ \mu$W) and $\varpi_2=-4.8$ dBm ($311\ \mu$W) to model the non-linear EH circuit, likewise the conversor in \cite{lopez_statistical_2019}. The mid-point of the linear region of this terminal's circuit amounts to $168.7\ \mu$W ($-7.73$ dBm).

The results shown in Fig. \ref{fig:distributions} are for a situation in which the pathloss, $\beta$, is such that the mean available power at each terminal is precisely $168.7\ \mu$W, which for a set of 24 terminals corresponds to a system's sum-power $P=4.05$ mW (6.07 dBm).
With the system's parameters summarized in Table \ref{tab:1_parameters}, the precoding gain achieved by the proposed optimal precoder can be calculated by using the link budget equation, such that: $G_P=6.07 \text{ dBm} -40 \text{ dBm} -10 \text{ dB} -\beta_{[\text{dB}]} = 19.57 \text{ dB}$. The chosen pathloss set to $\beta_{[dB]}=-63.5 \text{dB}$ corresponds to a $\sim23$ meters distance between the BC and the center of the clusters when using the LOS Urban Microcell 3GPPP propagation model at 1900 MHz \cite{Athanasiadou_06,3GPPmodel}), which is conceived for outdoor microcells, better matching the scenario expected for the proposed techniques, as sketched in Fig. \ref{Fig:system_setup}. For an indoor scenario one could consider the High Throughput Task Group (TGn) model D, that covers open spaces indoors environments with LOS \cite[p. 7]{Added_c_TGn}. In that case, a $\beta_{[dB]}=-63.5 \text{dB}$ would correspond to a 19.3 meters range.

One should note that the distributions in Fig. \ref{fig:distributions} are for the power available at the terminals. The effective harvested energy will have to consider the non-linear effects of \eqref{conversor} and the conversion factor $\eta$. Given the symmetric shapes of the sum-power distributions when precoders are used, the mean collected power at the $l$-th cluster is $\approx \eta P_l$. With $\eta=0.25$ ($-6$ dB) \cite{Mayaram2008,lopez_statistical_2019}, this leads to a mean harvested sum-power of $\approx 1$ mW by the 24 terminals.

Let us now analyze the limits of the available power obtained when using the unconstrained KL-based in the same analyzed in Fig. \ref{fig:distributions}. Considering the approximation of a sum-power distributed between 3.5 mW and 4.75 mW, the incoming power at each one of the 24 terminals is distributed between 148.8 $\mu W$ and 197.9 $\mu W$, which is well within the $[-22, \hspace{0.3em} -4.8]$ dBm interval, which corresponds to a $[6.3 \hspace{0.5em} 331.1] \mu W$ interval. 

\textit{Remark}: this set of parameters used in this subsection will later be referred to as the operation point (OP) in Fig. 2, and is summarized in Table \ref{tab:1_parameters}.

\subsection{Sum-power distributions: analysis}

The first thing worth mentioning is that when there are no restrictions to the minimum and maximum values of  $\mathbb{E} ||y_l(k)||^2$ (i.e., in the case of \textit{ideal} harvesting circuits), the results attained by the analytically designed precoders (given by (\ref{xl_optimal}) or (\ref{x_optimal}) and the precoders that result from the numerical solution of the linear programming problem with non-linear inequality constraints in \eqref{opt2} are exactly the same; they not only lead to the same  average available sum-power, $P$, in a multi-cluster or single-cluster system, but also exhibit the same PDF for the sum-power.
Interestingly, the particular solutions for the broadcast vectors $\mathbf{x}^*$ coming from the two approaches are often different, and therefore the two precoders $\mathbf{P}$ are different. However, these two different vectors $\mathbf{x}^*$ are equivalent in the sense that they lead to the same distributions for the same set of correlations $\mathbf{R}_l$ (and therefore for the same $\mathbf{A}$, defined in \eqref{opt_multi_clusters}. This is a consequence of the non-convexity of the optimization problem, exhibiting several equivalent local minima.

A paramount observation is that the sum-power distribution over the whole system obtained by the proposed technique, based on statistical CSIT, leads to a quasi-optimal energy harvesting situation, exhibiting a negligible loss with respect to the harvested sum-power when full-CSIT is available and the optimal precoder is used.
While the precoder based on the KL decomposition is quasi-optimal in respect to the full-CSIT case, by only setting the standalone goal of maximizing the sum-power available at the terminals may lead (for the KL-based or full-CSIT precoders) to a rather unbalanced distribution of the available power at the different clusters. In order to observe that effect, Fig. \ref{fig:distributions} also includes the distributions of the partial sum-powers at each of the three clusters, striking out that the average partial sum-power available at one of the clusters is one order of magnitude lower than the one available at the other two clusters. As seen in  Fig. \ref{fig:distributions}, two of the clusters hold very similar distributions, with mean values of 1.87 mW and 1.95 mW for their partial sum-powers, while the remaining cluster only gets a mean sum-power of 0.23 mW available at its terminals.

A fairness criterion for power allocation should take into account the effective power that the terminals harvest with the quasi-linear EH circuit characterized in \eqref{conversor}. This was framed in  \eqref{opt2}, such that the numerically obtained precoder attains a fair distribution of energy among the clusters in terms of mean partial sum-power. Fig. \ref{fig:distributions} also shows that the proposed constrained precoder attains the sought fairness, however at the expense of a reduction of the system's total sum-power, exhibiting the typical trade-off that exists in optimization problems when some fairness criteria is applied \cite{Thudugalage.2016},\cite{nguyen_energy_2019, nguyen_distributed_2017}. Notably, the system's sum-power obtained with numerical constrained precoding incurs an affordable degradation of $4.13-3.77=0.36$ mW ($8.7\%$) in respect to the average attained in the full-CSIT situation (the best possible one). 

For comparison purposes, two CSIT-free WET techniques that make use of linear arrays, namely the AA and SA schemes \cite{lopez_statistical_2019}, were also simulated and the results are also plotted in Fig. \ref{fig:distributions}. While the AA scheme provides a simple solution for spacial diversity, highly dependent on the angular deviation from the antenna's boresight, the SA is able to offer the same power for any angular position of a single cluster.
When using the AA technique, the power harvested by the \textit{k}-th terminal within the \textit{L}-th cluster is
\begin{equation}
P_{k_{\mathrm{AA}}}=\Omega\left(\frac{\beta(l)}{M}\left|\sum_{i=1}^{M} h_{i, k}\right|^{2}\right),
\end{equation}
while for the SA scheme that power is
\begin{equation}
P_{k_\mathrm{SA}}=\frac{1}{M} \sum_{i=1}^{M} \Omega\left(\beta(l)\left|h_{i,k}\right|^{2}\right).
\end{equation}
As previously stated, this paper focuses on the power available for EH, focusing on the RF power available at the antennas before considering the effect of $\Omega$.

\subsection{Validation with the baseline reference case}

\begin{figure}[t]
\centering
\subfigure[With uncorrelated Rayleigh fading, without LOS ($\kappa=0$). (Four curves overlap.)]
{
\includegraphics[width=0.95\columnwidth]{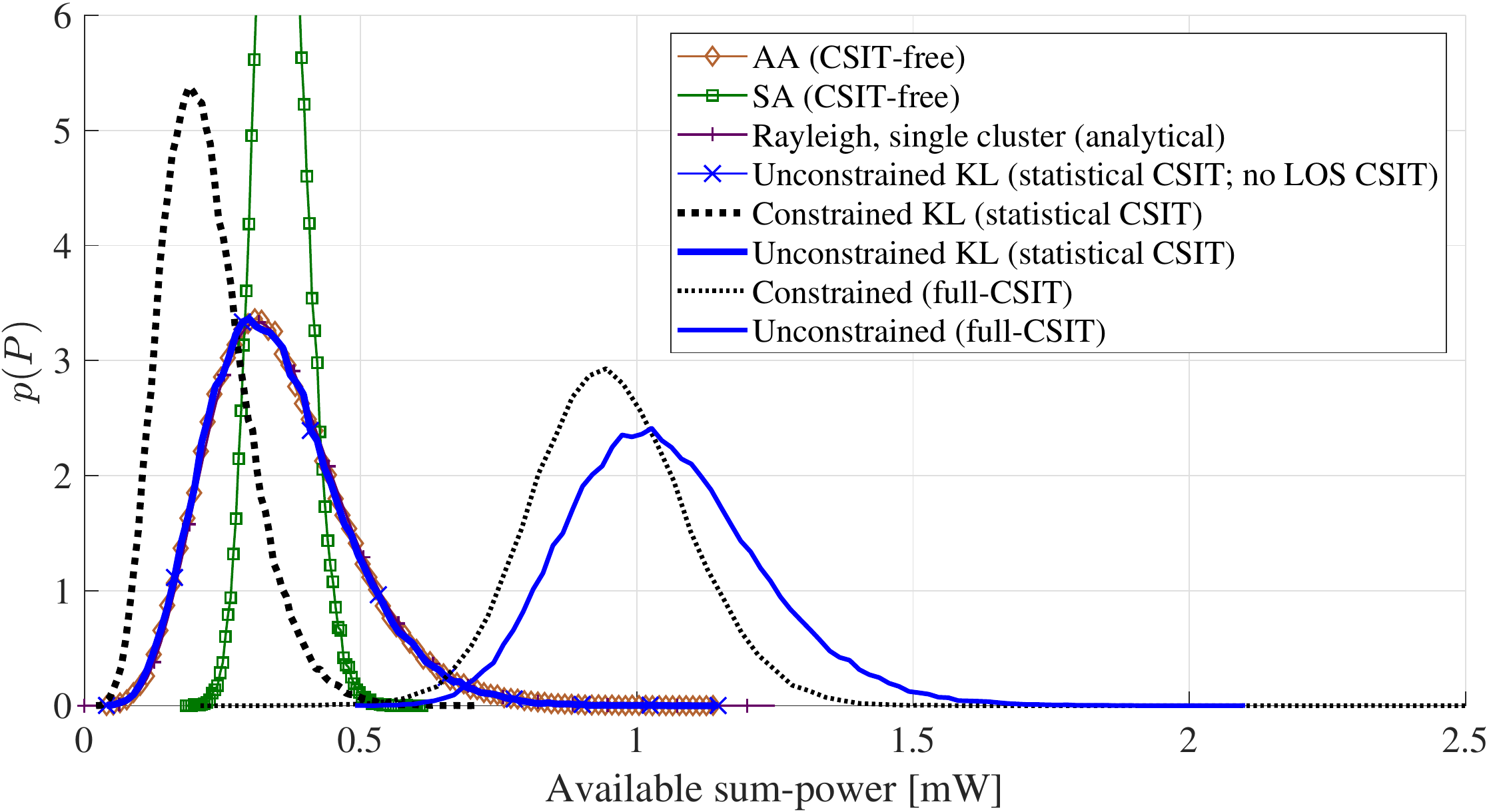}
}
\subfigure[With correlated Rayleigh fading, without LOS ($\kappa=0$). (Two curves overlap.)]
{\includegraphics[width=0.95\columnwidth]{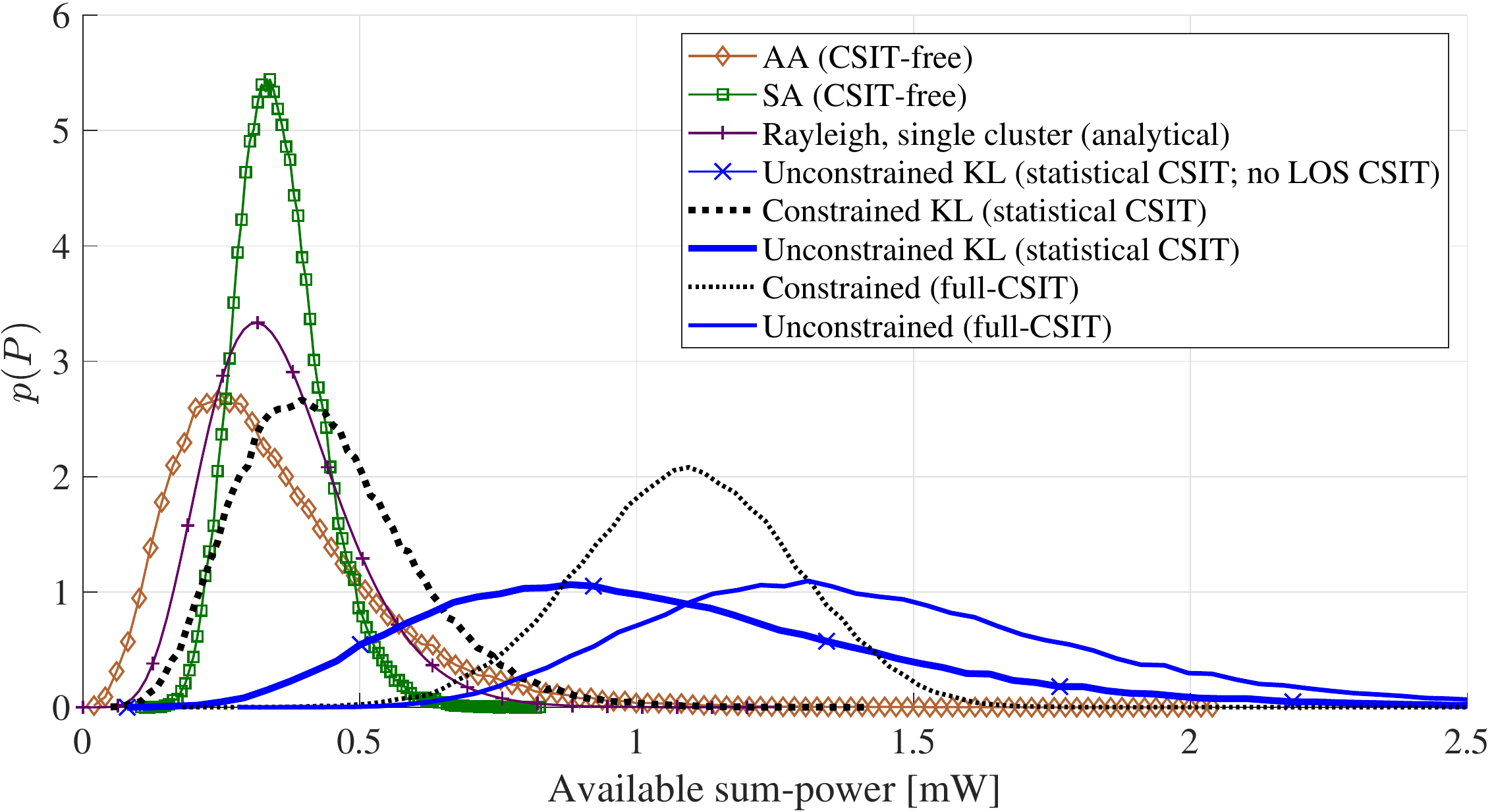}
}
\subfigure[With correlated Rayleigh fading, with LOS ($\kappa=1$)]
{
\includegraphics[width=0.95\columnwidth]{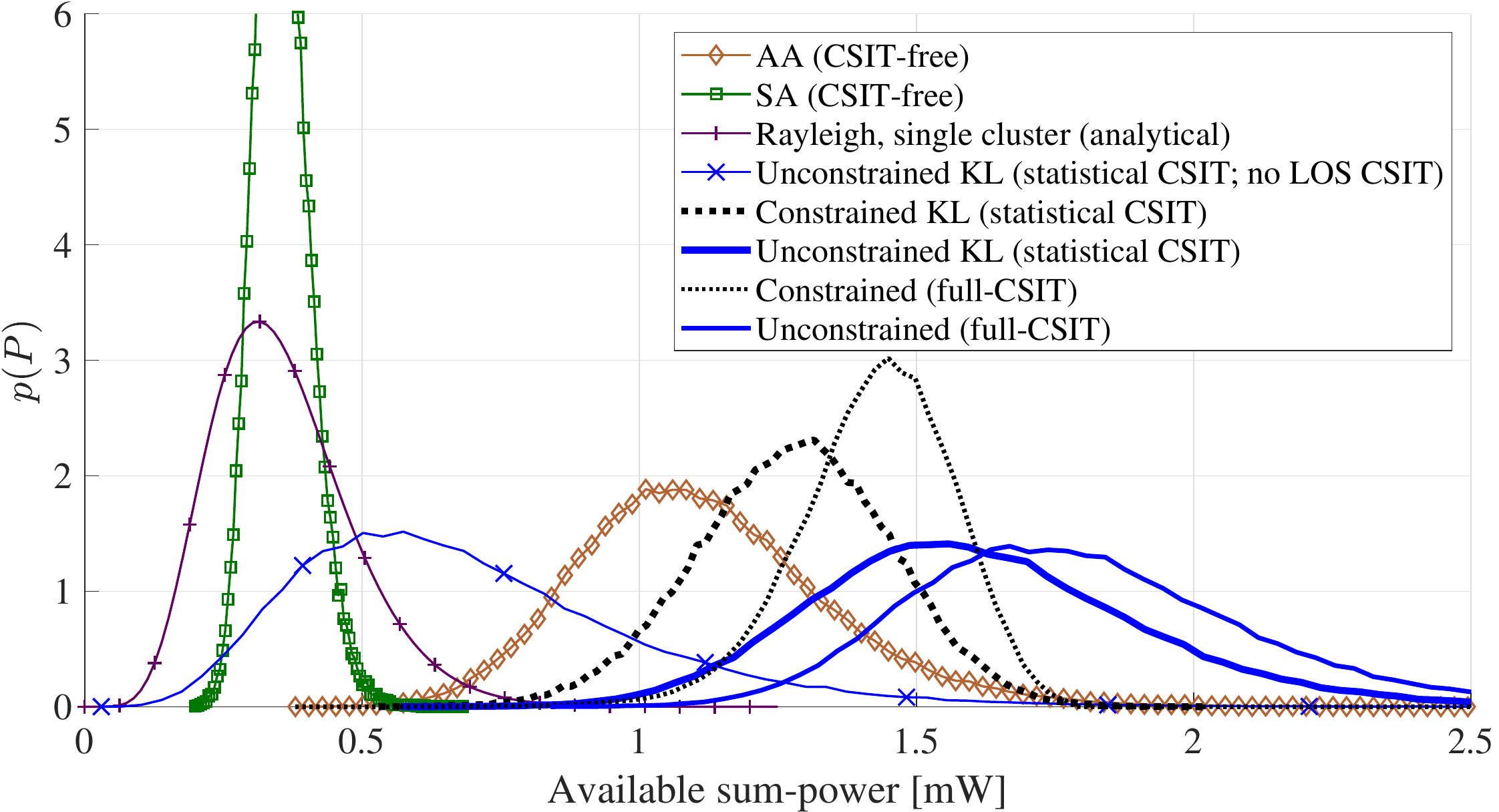}
}
\caption{Sum-power distribution for different precoding schemes ($M=8$ antennas, $K=8$ users per cluster) with $L=1$ cluster located at $\phi=85 \text{ degrees}$, including the analytical curve for the baseline reference case.}
\label{fig:distributions_gamma}
\end{figure}

The simulation of the analytical and numerical precoders was validated with the reference case where no clusters are considered and the link from the BS to all EH devices is an uncorrelated Rayleigh channel, as defined in section \ref{Sec: baseline}. This can be considered as the baseline situation with which the mean sum-powers attained by the different techniques can be compared to. Such results are plotted in Fig. \ref{fig:distributions_gamma}(a). It should be noted that the results with the AA scheme and with two of the precoded systems overlap the analytical curve defined by expression \eqref{eq: AA_gamma_sum-power} for the baseline reference case (with $KL=8$ terminals); these are i) the precoder derived for a channel with correlated Rayleigh and no LOS, as defined in \eqref{opt_MP}, and ii) the precoder derived for the Rician channel with a correlated Rayleigh component (but where now $\kappa=0$), as defined in \eqref{xl_optimal}. In this situation there is neither a notion of cluster nor of angle and the AA scheme performs as well as any of the precoded schemes. In this case, as also concluded in \cite{IoTJ_2021,lopez_statistical_2019}, the SA scheme is preferable because it provides a slightly larger mean sum-power and a significantly smaller variance of the PDF.

A case where there still no LOS component but the Rayleigh fading is correlated is shown in Fig. \ref{fig:distributions_gamma}(b): the optimal precoder for correlated fading, given by \eqref{opt_MP}, already exhibits the considerable gain that one can extract from the KL-based approach. In this case the PDF obtained using the precoder in \eqref{xl_optimal}, which also takes in consideration the LOS component, overlaps the previous curve, as expected, given that $\kappa=0$. However, when some LOS component emerges, as in the case in Fig. \ref{fig:distributions_gamma}(b), then the precoder for Rayleigh correlated fading has a subpar performance with respect the one incorporating information about the LOS component; as seen in Fig. \ref{fig:distributions_gamma}(c), the performance of the latter precoder detaches itself and an extra gain kicks in due to the extra information about the channel associated to the LOS CSIT (this gain can also be later seen in Fig. \ref{fig:evolution_Kappa} when looking at the $\kappa=0$ abscissa).

In this baseline case with uncorrelated fading in Fig. \ref{fig:distributions_gamma}(a), while there is no gain to be attained by the KL-based analytical precoders, the constrained numerical version of the proposed precoder (in sec. \ref{sec_numerical_opt}) narrows the sum-power distribution at the expense of a lower mean, as typical of fairness versus mean trade-offs. The same effect is narrowing the variance and reducing the mean also appears in the results of the constrained precoder in the situation of full-CSIT availability (the two curves with larger means in Fig. \ref{fig:distributions_gamma}(a)).

\textit{Remark}: the results presented in Fig. \ref{fig:distributions} and Fig. \ref{fig:distributions_gamma}, as well as the ones presented in the subsequent subsections, are obtained with $10^5$ simulated channels and are ergodic in the sense that for each channel instance a new fading correlation was generated.

\subsection{Rician factor}

\begin{figure}[tb]
\centering
\includegraphics[width=0.95\columnwidth]{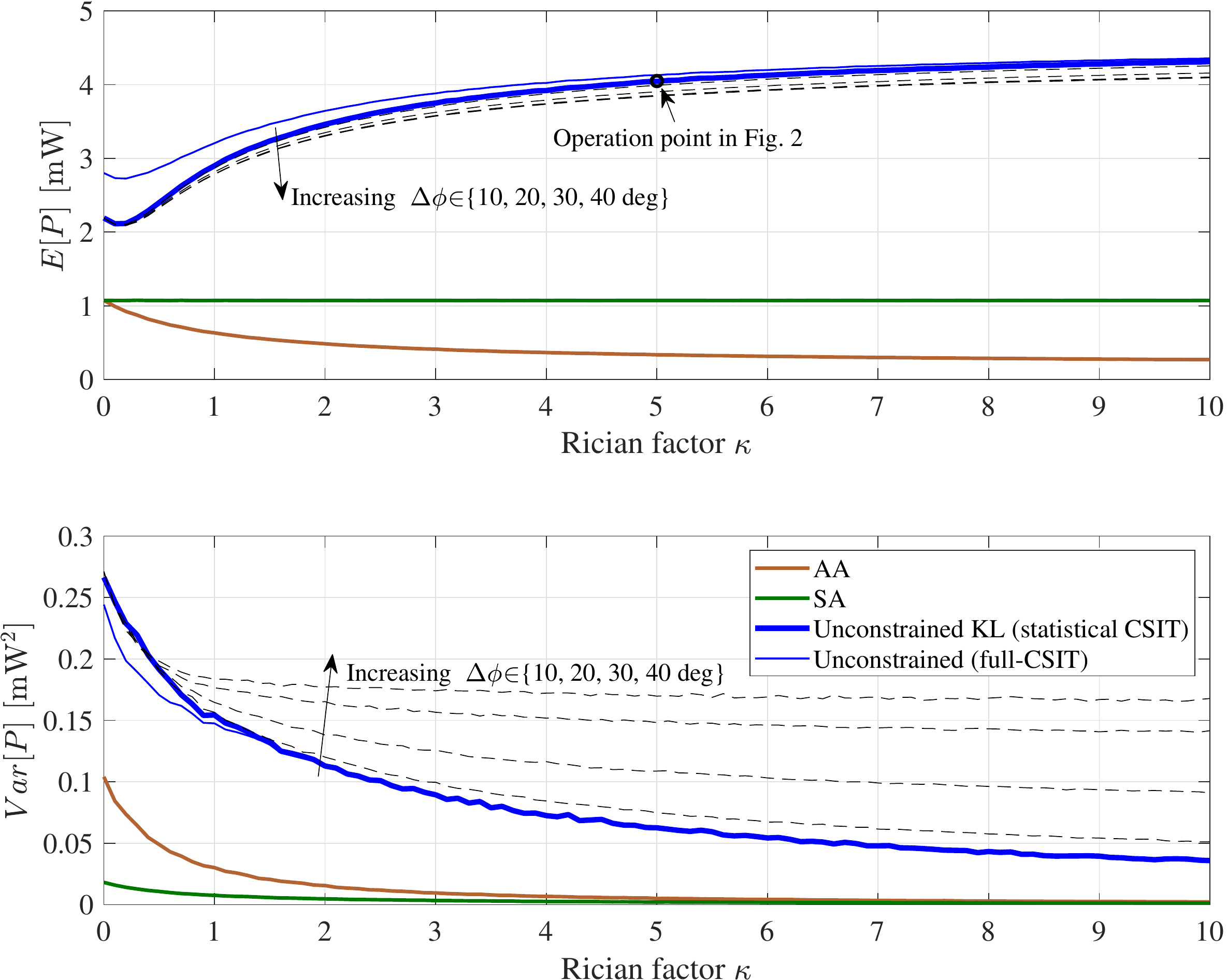}
\caption{Evolution of the mean and the variance of the sum-power for different precoding schemes for a varying Rician factor $\kappa$ (with $M=8$ antennas $K=8$ users) with $L=3$ clusters located at $\phi=\{0, 30, 70\} \text{ degrees}$. The black dashed lines show the effect of having the $K$ terminals within each cluster uniformly distributed over an angular aperture $\Delta_\phi$.}
\label{fig:evolution_Kappa}
\end{figure}
The rich information provided by how the different PDFs vary when the system's parameters change motivates a more detailed study of how each of the system's parameters impact on the sum-power distributions, specifically how their mean and variance is modified. The impact of the LOS component will be the first to be analyzed in Fig. \ref{fig:evolution_Kappa}, while maintaining all the parameters on Table \ref{tab:1_parameters} unchanged.

When the LOS component grows larger in comparison to the Rayleigh fading component, the channel becomes more deterministic and for that reason the performance tends to the one achieved with full channel knowledge. One should notice in Fig. \ref{fig:evolution_Kappa} that the OP scrutinized in Fig. \ref{fig:distributions} already corresponds to a regime in which CSIT leads to a very diminished increase in $\mathbb{E}(P)$, unlike what happens in the range where $\kappa<1$ (even though in that region the power spreads over a wider domain, as depicted in the curve for $Var[P]$).

The advantage of the proposed precoding scheme stands out in comparison with the energy harvesting ability of both the SA and AA schemes. Noticeably, at the abscissa $\kappa=0$ it is possible to see an over 3 dB gain. In this situation, without a LOS component, all the gain appears by virtue of the KL-based approach, by having access to the second orders statistics of the fading only. Then, an increasing weight of the LOS increased the mean sum-power while simultaneously diminishing its variance, given that the channel tends to become deterministic. Fig. \ref{fig:evolution_Kappa} also shows the effect of considering the terminals uniformly scattered over an increasing angular aperture $\Delta_\phi$ around the central angle of each cluster.

\subsection{Array size}

\begin{figure}[tb]
\centering
\subfigure[With $M=2^n$, for $M=1$ to 128]
{
\includegraphics[width=0.95\columnwidth]{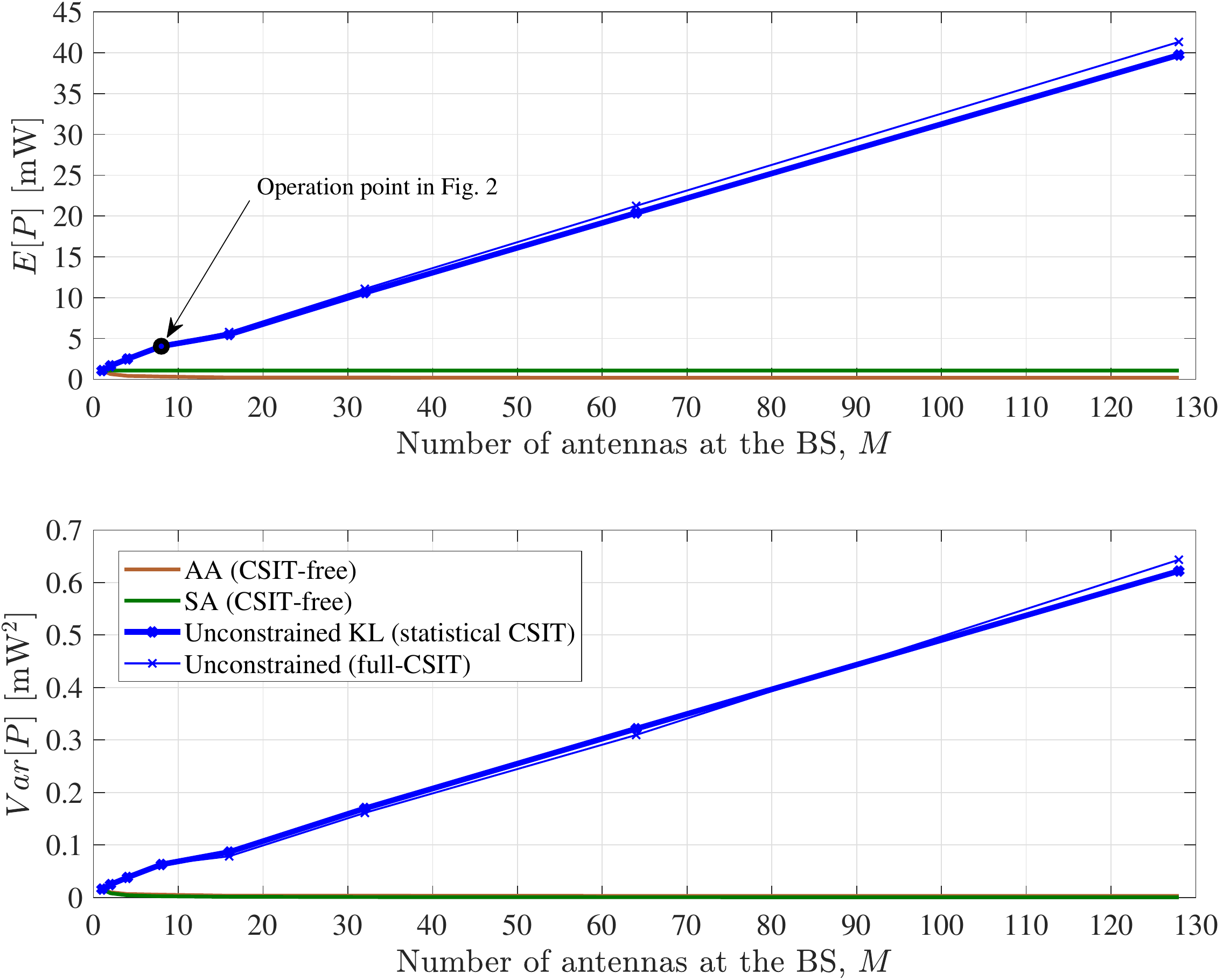}
}
\subfigure[With an incremental unit step for $M=1$ to 128]
{\includegraphics[width=0.95\columnwidth]{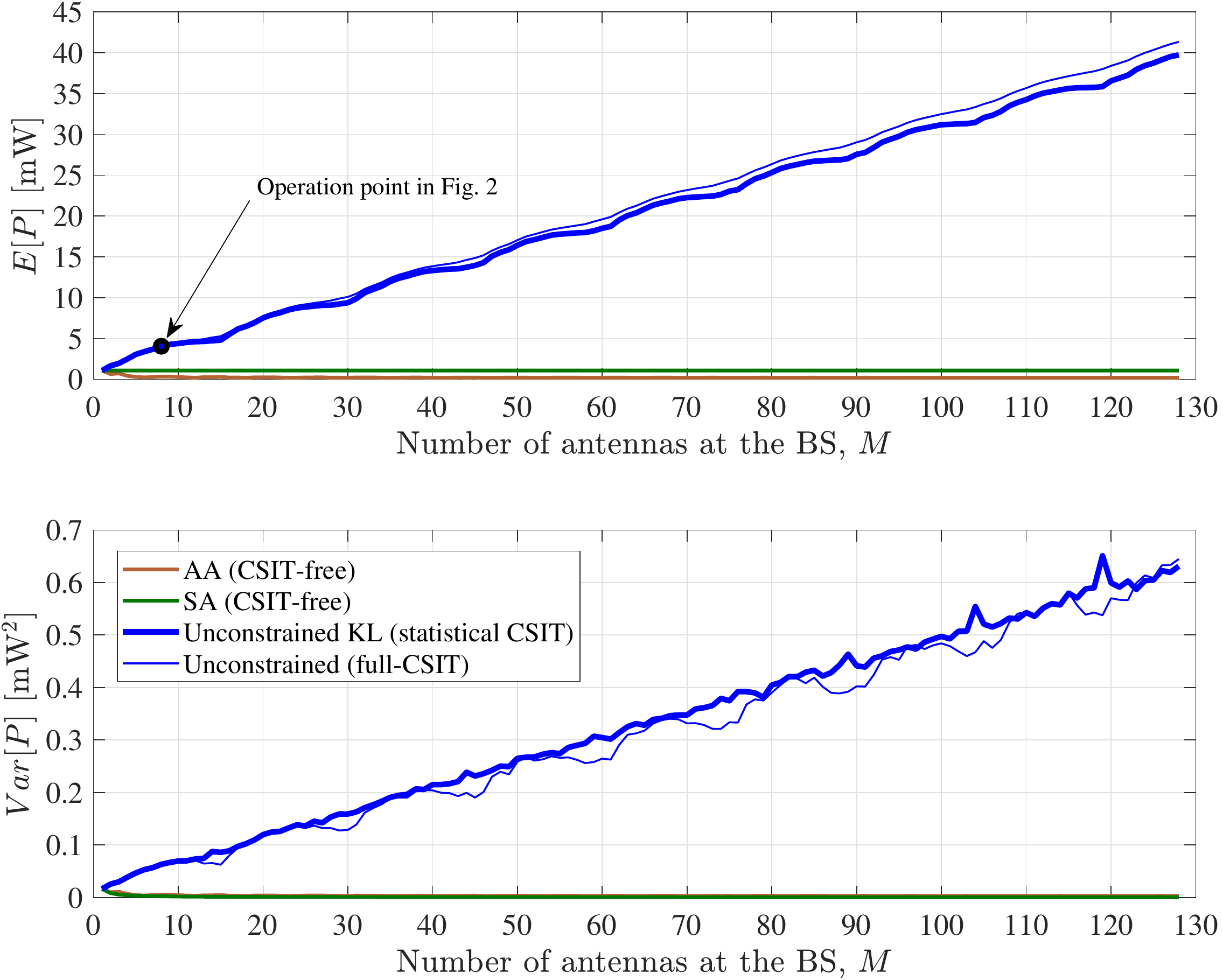}
}
\caption{Evolution of the mean and the variance of the sum-power for different precoding schemes for a varying number of antennas, $M$, at the BS with $L=3$ clusters located at $\phi=\{0, 30, 70\} \text{ degrees}$ (with $K=8$ users per cluster, and $\kappa=5$).}
\label{fig:evolution_M}
\end{figure}

When the BS is fitted with a larger number of antennas, $M$, one can observe in Fig. \ref{fig:evolution_M}(a) a quasi-linear growth of the system's available power for harvesting when the ULA's number of antennas is a power of two (as it is typical in commercial ULAs). The increase in $\mathbb{E}[P]$ comes in tandem with a quasi-linear spread of the variance, meaning that higher gains are intertwined with a less constant sum-power level. This increasing focusing effect onto the clusters, translates to larger precoding gains for WET that are unattainable by AA (apart when aligned with the antenna's boresight) or even SA.
For the particular set of three clusters tested in Fig. \ref{fig:evolution_M}, an ULA with $M=128$ antennas attains a 40 fold gain (16 dB) in respect to the AA scheme, and this is before considering any optimization of the ULA's rotation, as will be seen in the next subsection.

It should be noted that, while in the case with a single cluster both the mean and the variance always exhibit a linear growth (not shown in the paper) for any step size of the incremental $M$, when $L>1$ the variance drops bellow linear growth in a periodic manner when analyzing the results for an ULA growing one antenna at a time, as can be observed in Fig. \ref{fig:evolution_M}(b). It has been found that this is due to the radiation pattern and the particular positions of the $L$ clusters in respect to the radiation pattern.

\begin{figure}[t]
\centering
\includegraphics[width=0.95\columnwidth]{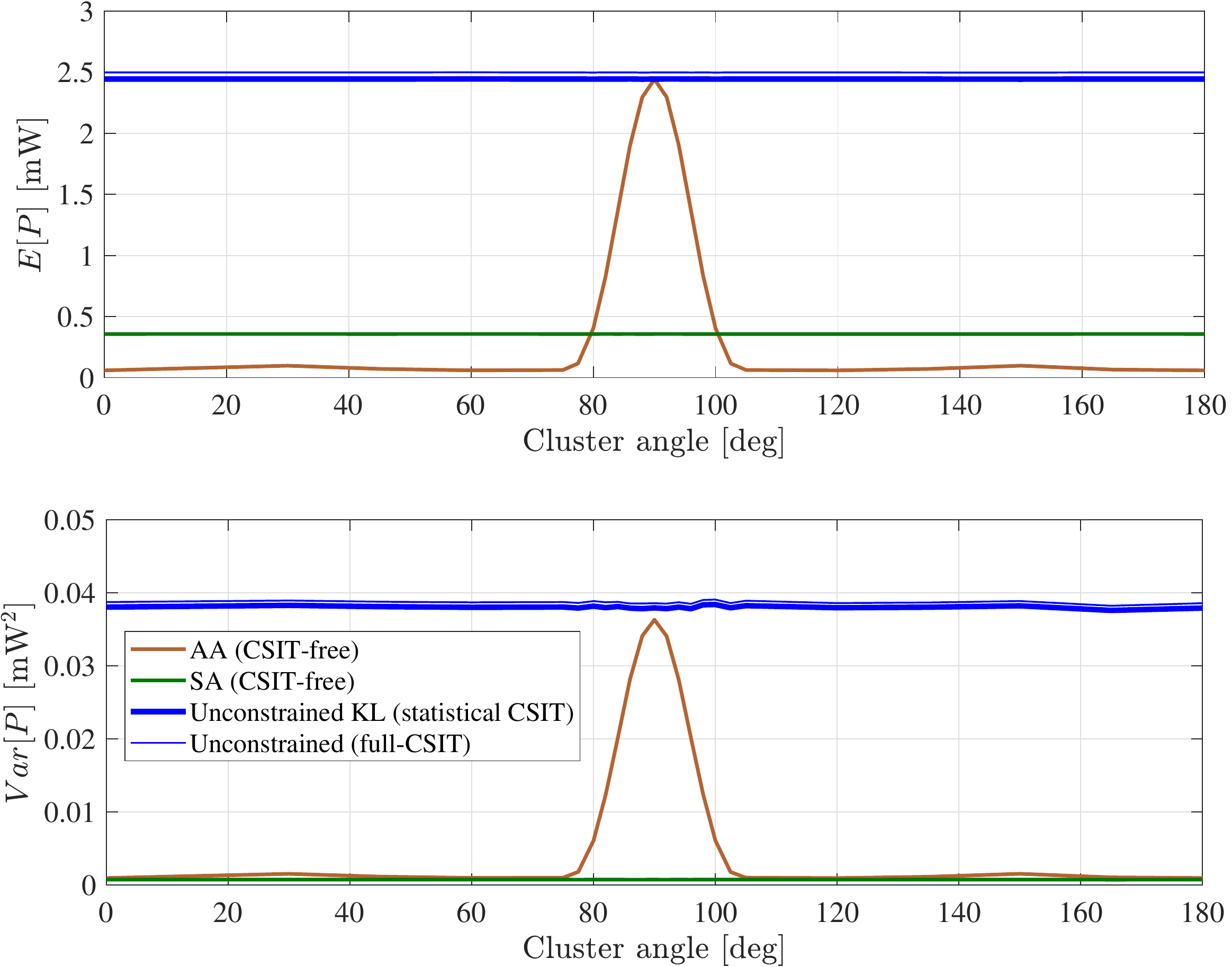}
\caption{Evolution of the mean and the variance of the sum-power in the case of a single cluster ($L=1$) positioned at different angles (with $M=8$, $K=8$ users, and $\kappa=5$).}
\label{fig:evolution_phi}
\end{figure}

\begin{figure}[t]
\centering
\includegraphics[width=0.95\columnwidth]{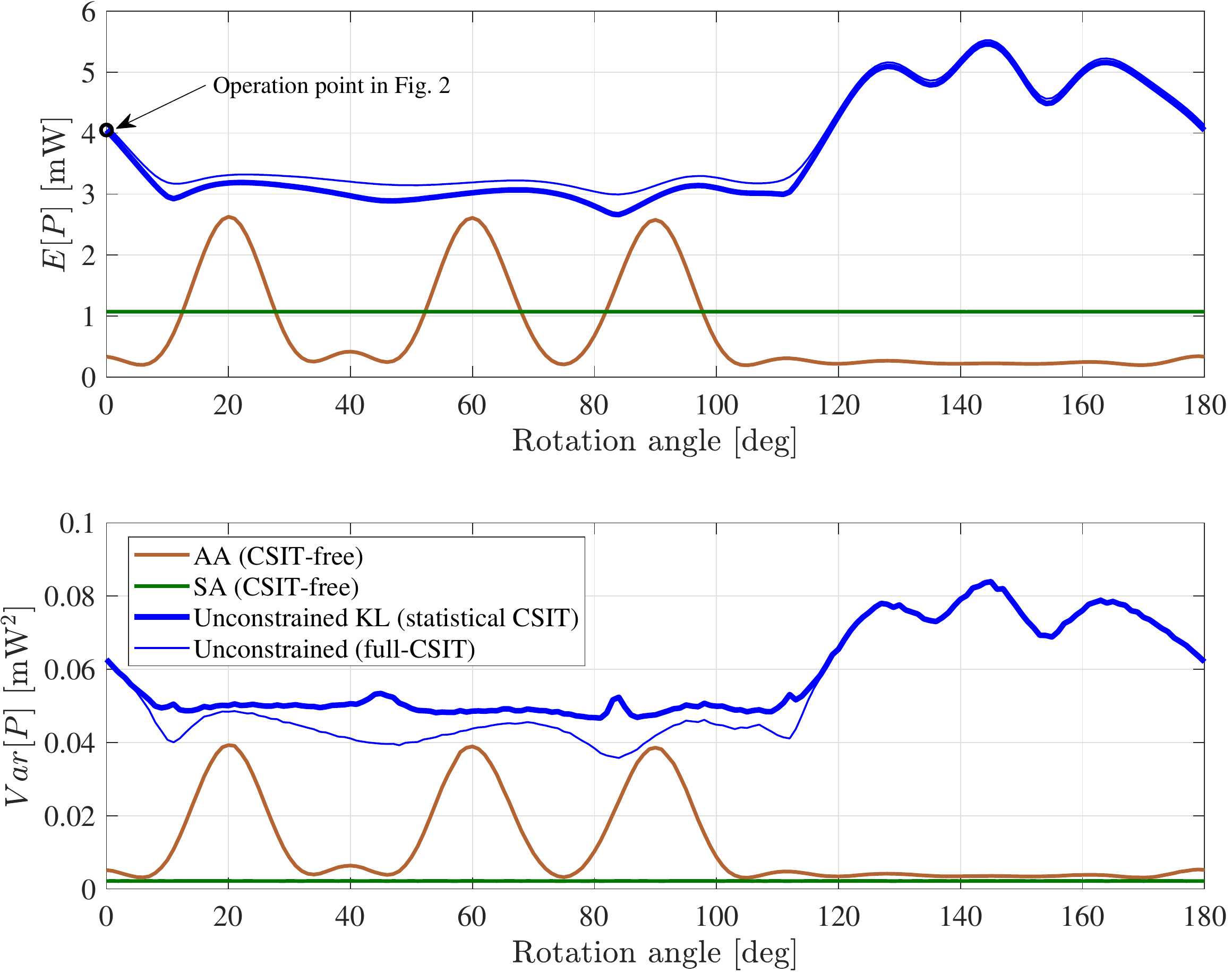}
\caption{Evolution of the mean and the variance of the sum-power for a varying ULA's rotation angle $\psi$, with $L=3$ clusters located at $\phi=\{0, 30, 70\}$ degrees (when $\psi=0$) (with $M=8$, $K=8$ users per cluster, and $\kappa=5$).}
\label{fig:evolution_rotation}
\end{figure}

\subsection{Cluster's angular position}

In order to analyze the effect of the angular cluster's positions,  Fig. \ref{fig:evolution_phi} depicts a situation of a single cluster positioned at varying angles. It is known (e.g., \cite{lopez_statistical_2019}) that the AA scheme can provide a significant gain due to constructive interference but limited to the ULA's boresight. The SA scheme can provide a constant mean sum-power for any angle, however at a much lower level than AA's peak. Fig. \ref{fig:evolution_phi} shows that the KL-based precoder can sustain the AA's peak at all angles, and again, in the considered scenario one finds a negligible loss with respect to the mean sum-power attained with full-CSIT.

The angular positions of the clusters are relative to the geometry of the array, i.e., relative to the broadside and to the end-fire of the ULA. The $l$-th cluster is considered to be positioned at angle $\phi_l$ in respect to the ULA's end-fire, such that an array facing the boresight has and angular position $\phi_l=90$ degs. If the linear array can be mechanically rotated by a certain angle $\psi$, then the effective new positions of a set of $L=3$ clusters change to
\begin{equation}
\Phi=\phi+\psi=\{ \phi_1, \phi_2, \phi_3 \}+\psi,
\quad \psi \in \left[ 0, \pi \right]
\end{equation}
In Fig. \ref{fig:evolution_rotation} one assesses the effect of such rotation starting with $\phi=\{0, 30, 70\}$ degrees, the set previously used in Fig. \ref{fig:distributions}. Given the single-lobe directivity of the AA technique (as seen in Fig. \ref{fig:evolution_phi}), one finds peaks for the sum-power when the new effective positions of each cluster is 90 degrees, which for $\phi=\{0, 30, 70\}$ degrees, happens for $\psi=20,60$, and 90 degrees. Notably, one finds that there is an optimal rotation that maximizes the system's sum-power for $\psi=144$ degrees, attaining $\mathbb E\{P\}=5.458$ mW. For this particular positions of the clusters, the proposed unconstrained scheme attains the same sum-power mean and variance of the full-CSIT case, for a wide domain of rotations $\psi$. The need for optimizing the ULA's rotation in WET has also been noted in \cite{WCL,IoTJ_2021B}.

\subsection{Number of clusters}

\begin{figure}[t]
\centering 
\includegraphics[width=0.95\columnwidth]{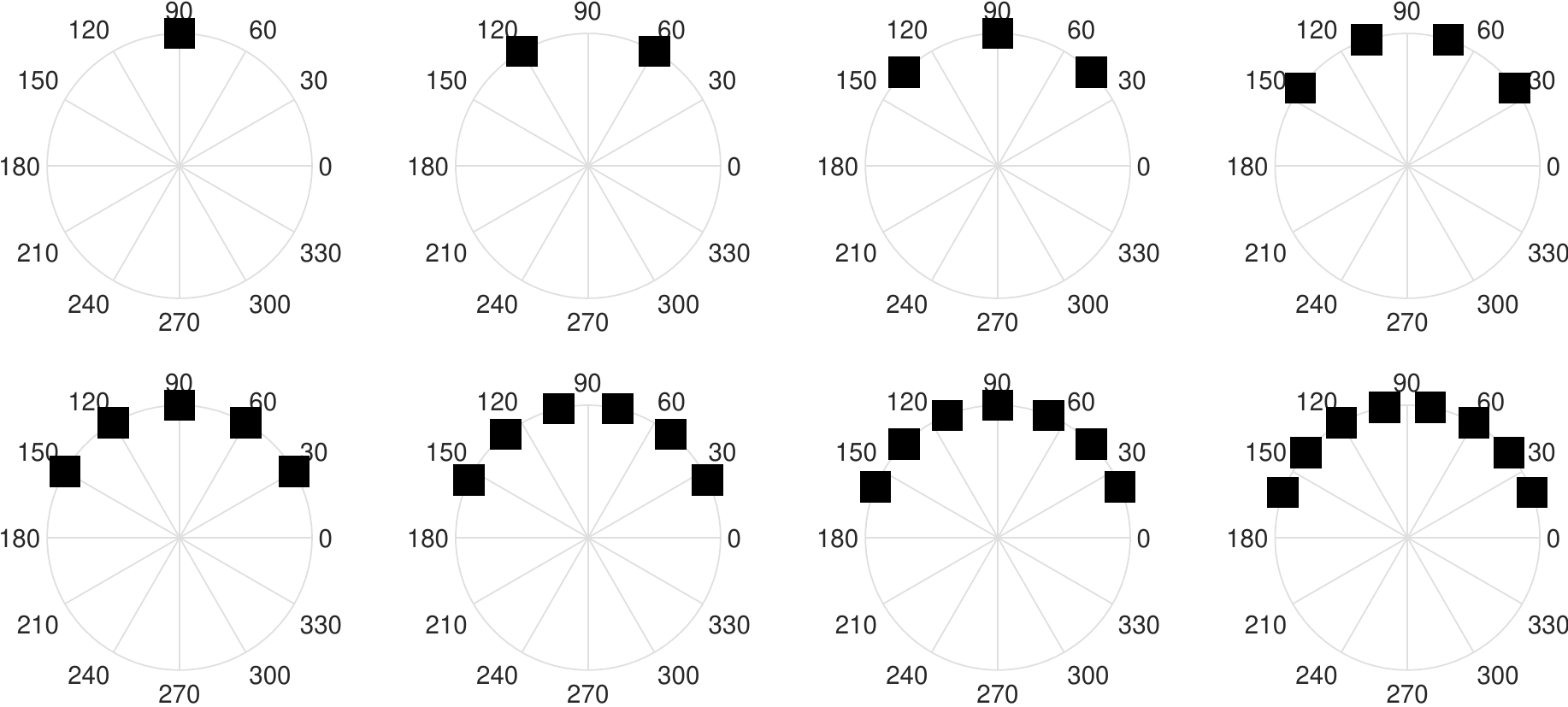}
\caption{Angular positions of the clusters from $L=1$ to $L=8$.}
\label{Fig: clusters_positions}
\end{figure}

\begin{figure}[t]
\centering 
\includegraphics[width=0.95\columnwidth]{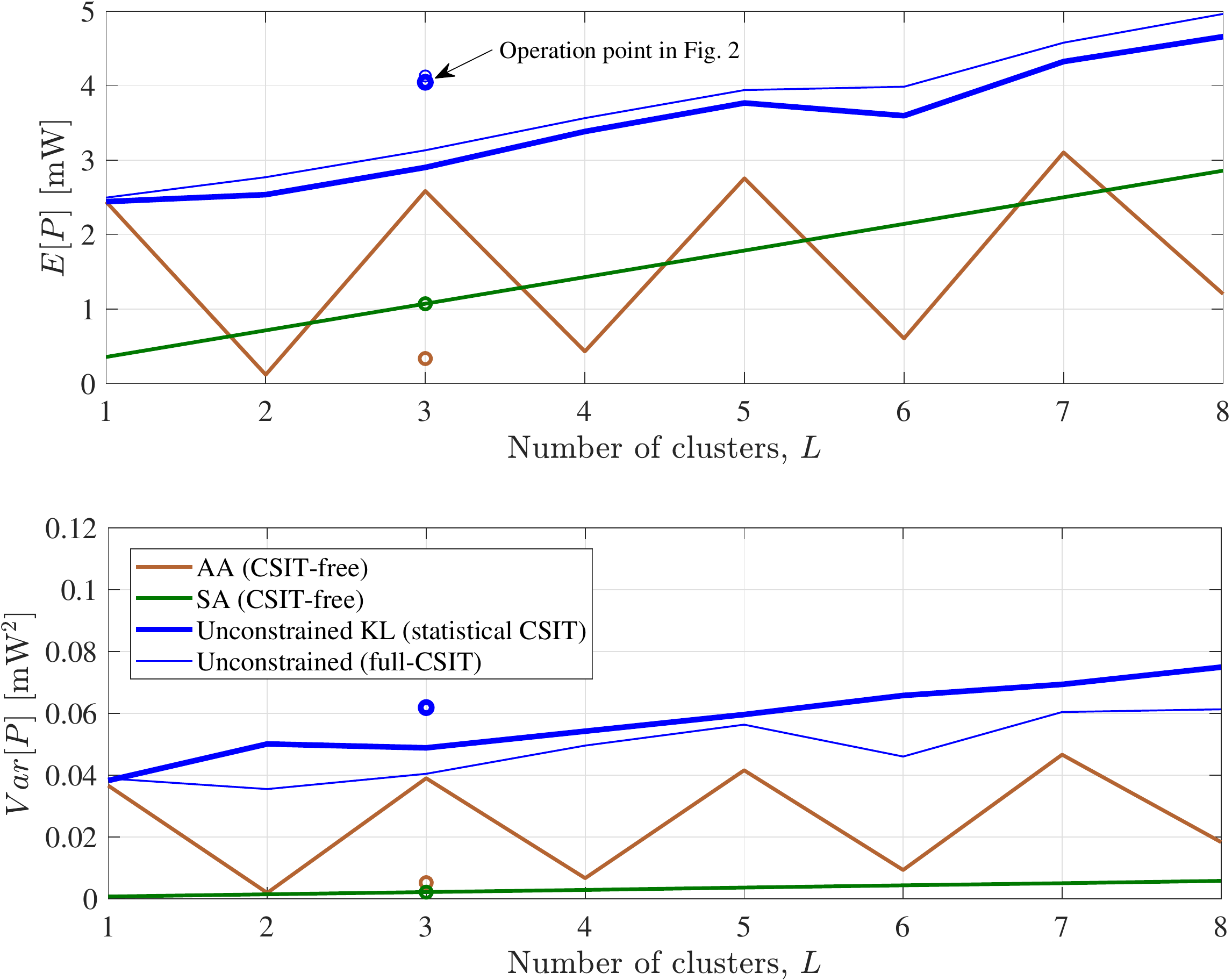}
\caption{Evolution of the mean and the variance of the sum-power for a varying number of clusters, $L$, (with $M=8$, $K=8$ users per cluster, and $\kappa=5$). The angles of the $L$ clusters are given by \eqref{clusters}. The case with $L=3$ clusters located at $\phi=\{0, 30, 70\}$ is also plotted with marks (in the case of the variance the two blue marks fully overlap).}
\label{Fig: evolution_L}
\end{figure}

With a constant number of $K$ devices per cluster, an increasing number of clusters $L$ leads to a proportional increase of the number of terminals in the system and  $\mathbb{E}\{P\}$ increases as a direct consequence of having more harvesting terminals in the system. The simulation results consider clusters positioned at the angles 

\begin{equation} \label{clusters}
 \phi_l=\frac{l}{L+1}\pi, \quad \text{for} \quad l=1,2,\ldots, L,
\end{equation} 

\noindent as depicted in Fig. \ref{Fig: clusters_positions} up to eight clusters, and the corresponding results are plotted in Fig. \ref{Fig: evolution_L}. This choice of angular symmetries leads to an apparent non-monotonicity of the harvested energy in the case of AA system. This is illusory given that, for an odd number of clusters, \eqref{clusters} places a central cluster at the boresight of the ULA, which, as seen in the previous subsection in  Fig. \ref{fig:evolution_phi}, is the direction of a narrow beam that chiefly powers that central cluster. In fact, both the maxima (at odd $L$) and the minima (at even $L$) also increase with $L$, as more clusters are packed together and the ones neighboring the central cluster increasingly get more illuminated by the AA's beam.

The values obtained with the previous setup, with three clusters located at $\phi=\{0, 30, 70\}$ degrees, are also superimposed on Fig. \ref{Fig: clusters_positions} (at abscissa $L=3$ ), and one can observe that that set of angles leads to a larger mean available power, with larger variance, then the set stipulated by \eqref{clusters}.

\section{Discussion and Conclusions}
In the context of mMTC, WET is a solution for BS wirelessly power a very large number of EH devices.
However, in many practical situations the terminals may appear clustered (due to physical proximity), and their channels to the BS may have several parameters in common.
Considering Rician channels from the BS to the terminals, the terminals within defined clusters were assumed to share a common correlation of their Rayleigh fading components, as well as having a slowly varying LOS channel component (possibly different but fully-known at the BS) and sharing the same large-scale pathloss.
Rather than considering the partial channel knowledge of the non-LOS fading component as the sum of a known and an unknown term, this paper considers a statistical representation of that random component.
The proposal takes advantage of the Karhunen–Loève representation of a MIMO channel, such that the correlated Rayleigh component can be written as the concatenation of i) an unknown fast-changing Rayleigh uncorrelated component and ii) CSIT only depending on the first and second order statistics of the channel (mean pathloss and correlation, respectively) for the coherence duration of the fading's autocorrelation. In doing so, the paper proposed beamforming precoders for multi-antenna BS in WET systems operating with low-rate statistical CSIT. These precoding schemes require low computational complexity, chiefly depending on SVD calculations for each one of the clusters. 

The results exhibit quasi-optimal performance under Rician fading when compared with the ones obtained when beamforming with full-CSIT knowledge, even for Rician factors in the interval $\kappa \in [1 \hspace{0.5em} 5]$ (i.e., below the one considered in \cite{WCL}), and, in particular, a positive gain exists even for $\kappa=0$, by just leveraging the existence of correlated fading.

Because the number of clusters is bound by the maximum number of beams, one has $L \leq M$ and therefore the use of a massive array is highly desirable. The precoding gain grows linearly with the number of antennas at the BS (at the expense of wider power spread) and therefore the desideratum of wirelessly power a large number of terminals without instantaneous CSIT can be fulfilled. The same linear increase of the mean sum-power also exists when using a precoder with access to instantaneous full-CSIT. The available CSIT-free multi-antenna schemes for WET have a much poorer performance in similar conditions and are not able to take full advantage of a larger array, nor to provide multiple energy beams to different sites.

By setting as the only goal the maximization of the sum-power in a multi-cluster, an unfair power allocation between the different clusters may emerge. To withstand that effect, a constrained optimization of the precoders is proposed which, besides naturally enforcing inter-cluster fairness, can also conform the power domain at the terminals' antennas to the linear domain of their non-linear EH circuit.
For the chosen setup with three clusters with 24 terminals in total, with a pathloss of 63.5 dB from the BS to each of the clusters, corresponding to $\sim23$ meters at 1900 MHz (by lowering the frequency the range will be larger), neither the saturation point of the EH non-linear circuit considered was ever reached, nor the power available per terminal fell below the EH sensitivity. In fact, the available power at each terminal's antenna is always within the  limits of the linear domain of the EH circuit.

In environments with a reduced number of propagation paths or with highly correlated paths, one could reduce the dimensionality of the problem by neglecting some of the singular values of the autocorrelation matrix, similarly to the idea in \cite{jelitto_reduced_2002}, which made use of the related Karhunen–Loève transformation \cite[App. E]{Haykin}.

Similarly to the conclusions in \cite{WCL}, where a mechanical rotation of the ULA is found to be a parameter to be optimized, the present research has shown that, in the case of multi-cluster systems, a similar mechanical rotation should also be optimized such that the clusters are placed in a set of more favorable angles. Interestingly, it was recently found in a context of LOS channels that an ULA can attain a performance very close to capacity with proper angular rotation, even without the need of a mechanical rotation, but rather by selecting a particular array out of a very low number of fixed ULAs with different relative rotations \cite{Lozano_reconfigurable_2020}.

As  an  immediate  future  work,  it  would  be  interesting  to  combine  the herein  proposed  beamforming optimization with the optimization of the waveform \cite{Clerckx.2018, added_a36, Added_d_MTTW} and analyze the reachable energy savings at the BS for the same EH constraints at the devices. This research can also be extended to the realm of SWIPT, with modulated waveforms \cite{Added_d_Clerckx}.
While statistical information can be more accurately obtained, the impact of working with imperfect statistical CSIT should also be evaluated.
Additionally, even though the proposed signal processing techniques for precoding are blind to the geometry of the antenna arrays at the BS, the linear arrays considered in this work only permit a 2D tuning of the radiation pattern. Hence, for certain deployment where the terminals are also spread in height (e.g., in some types of infrastructures monitored by sensors) it will be important to extend the techniques to MIMO beamforming with 3D steering capability \cite{OJVhT_2020}.
The constrained optimization problem can be generalized to the situation where the terminals have different sensitivity and saturation points. A likely scenario could be the one of having different types of equipment associated to different clusters, for example, monitoring different types of infrastructure.

\section*{Acknowledgments}
The authors are grateful to Prof. Petar Popovski (Aalborg University) for discussions about modeling of wireless EH systems, to Prof. Angel Lozano (Universitat Pompeu Fabra) for discussions on wireless propagation models and on constrained optimization, and to Prof. Ioannis Krikidis (University of Cyprus) for discussions on beamforming for WET. 

\bibliographystyle{IEEEtran}
\bibliography{Fairness_WirelessEnergTrans.bib,Fairness_WirelessEnergTrans_BOOKS.bib, reply_additions.bib}

\begin{IEEEbiography}[{\includegraphics[width=1in,height=1.25in,clip,keepaspectratio]{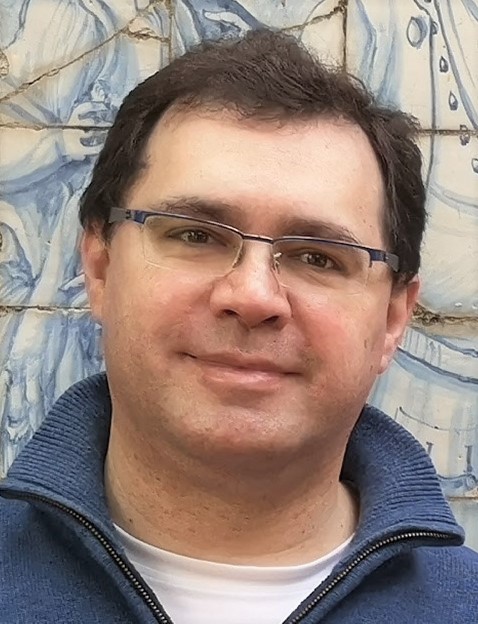}}]{Francisco A. Monteiro}
(S’07–M’13) is Assistant Professor in the Dep. of Information Science and Technology at Iscte - University Institute of Lisbon, and a researcher at Instituto de Telecomunicações, Portugal. He holds a PhD from the University of Cambridge, UK, and the Licenciatura and MSc degrees in ECE from IST, University of Lisbon. He held visiting research positions at the Universities of Toronto (Canada), Lancaster (UK), Oulu (Finland), and Pompeu Fabra (Barcelona, Spain). He got a number of best conference paper and exemplary reviewer awards from IEEE. He co-edited a book on MIMO, published by CRC in 2014, was the Lead Guest Editor of a special issue on Network Coding on EURASIP JASP in 2016, and was a general chair of ISWCS in 2018.
\end{IEEEbiography}

\begin{IEEEbiography}[{\includegraphics[width=1in,height=1.25in,clip,keepaspectratio]{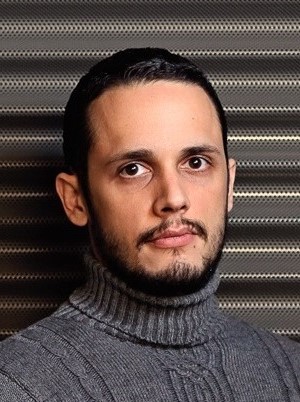}}]{Onel L. A. López}
(S’17–M’20)  received the B.Sc. (Honours), M.Sc. and D.Sc. (with Distinction) degrees in ECE from Central University of Las Villas (Cuba), Federal University of Paraná (Brazil), and University of Oulu (Finland), respectively.
During 2013-2015, he worked as telematics specialist in the Cuban telecommunications company (ETECSA).
He is a co-recipient of the 2019 EuCNC Best Student Paper Award and a collaborator to the 2016 Research Award given by the Cuban Academy of Sciences. He is now Assistant Professor in the Centre for Wireless Communications (CWC), Oulu, Finland.
\end{IEEEbiography}

\begin{IEEEbiography}[{\includegraphics[width=1in,height=1.25in,clip,keepaspectratio]{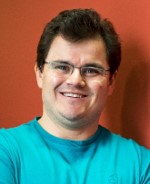}}]{Hirley Alves}
(S’11–M’15) is Assistant Professor and Head of the Machine-type Wireless Communications Group at the 6G Flagship, Centre for Wireless Communications, University of Oulu. He leads the MTC/URLLC activities for the 6G Flagship Program. He is actively working on massive connectivity and ultra-reliable low latency communications for future wireless networks, 5GB and 6G, full-duplex communications, and physical-layer security.  He was the General Chair of the ISWCS’2019 and the General Co-Chair of the 1st 6G Summit, Levi 2019, and ISWCS 2021.
\end{IEEEbiography}

\end{document}